\documentclass[twocolumn,english,aps,prl,showpacs,superscriptaddress,groupaddress,floatfix,graphics,graphicx]{revtex4-1}
\usepackage{lmodern}
\usepackage[T1]{fontenc}
\usepackage[latin9]{inputenc}
\usepackage{xcolor}
\usepackage{babel}
\usepackage{float}
\usepackage{amsmath}
\usepackage{amssymb}
\usepackage{graphicx}
\usepackage{setspace}
\usepackage[unicode=true,pdfusetitle,
 bookmarks=true,bookmarksnumbered=false,bookmarksopen=false,
 breaklinks=true,pdfborder={0 0 1},backref=false,colorlinks=false]
 {hyperref}
\hypersetup{
 colorlinks,linkcolor=blue,citecolor=blue,urlcolor=blue}

\makeatletter

\newcommand*\LyXThinSpace{\,\hspace{0pt}}
\providecommand{\tabularnewline}{\\}

\usepackage{babel}
\usepackage{babel}

\usepackage{blindtext}

\usepackage{dsfont}\usepackage{bbm}\IfFileExists{lmodern.sty}{\usepackage{lmodern}}{}
\usepackage{babel}

\makeatother

\begin{document}

\title{Anomalous Hall effect in 2D Dirac materials}

\author{Manuel Offidani }

\affiliation{Department of Physics, University of York, York YO10 5DD, United
Kingdom}

\author{Aires Ferreira}
\email{aires.ferreira@york.ac.uk}

\selectlanguage{english}%

\affiliation{Department of Physics, University of York, York YO10 5DD, United
Kingdom}
\begin{abstract}
We present a unified theory of charge carrier transport in 2D Dirac
systems with broken mirror inversion and time-reversal symmetries
(e.g.,\textcolor{black}{{} as realized in ferromagnetic graphene). }We
find that the entanglement between spin and pseudospin SU(2) degrees
of freedom stemming from spin\textendash orbit effects leads to a
distinctive gate voltage dependence (change of sign) of the anomalous
Hall conductivity approaching the topological gap, which remains robust
against impurity scattering and thus is a smoking gun for magnetized
2D Dirac fermions. Furthermore, we unveil a robust skew scattering
mechanism, modulated by the spin texture of the energy bands, which
causes a net spin accumulation at the sample boundaries even for spin-transparent
disorder. The newly unveiled extrinsic spin Hall effect is readily
tunable by a gate voltage and opens novel opportunities for the control
of spin currents in 2D ferromagnetic materials. 
\end{abstract}
\maketitle
Ferromagnetic order in two-dimensional (2D) crystals is of great significance
for fundamental studies and applications in spintronics. Recent experiments
have revealed that intrinsic ferromagnetism occurs in 2D crystals
of Cr$_{2}$Ge$_{2}$Te$_{6}$\,\cite{Cr2Ge2Te6_Gong_17} and CrI$_{3}$\,\cite{CrI3_Huang_17},
while graphene and group-VI dichalcogenide monolayers acquire large
exchange splitting when integrated with nanomagnets \cite{Ferr_G_YIG_Swartz12,Ferr_G_YIG_Wang15,AHE_G_Tang_18,Ferr_G_YIG_Leutnantsmeyer_17,Ferro_Gr_EuS_Wu17,Ferro_Gr_BiFeO3_Wu17,Ferro_vdW_mat}. 

Different from bulk compounds, the electronic states of atomically
thin layers can be dramatically affected by short-range magnetic interactions,
opening up a new arena for studies of emergent spin-dependent phenomena
\cite{MEC_Gr_Theory_Haugen08,MEC_Gr_Theory_QAHE_Sun_10,MEC_Gr_Theory_Sun11,MEC_Gr_Theory_Yang13,MEC_Gr_Theory_Marchenko15}.
In this regard, graphene and other 2D Dirac materials with multiple
internal degrees of freedom offer particularly  promising perspectives.
The anomalous Hall effect (AHE) recently observed in graphene/yttrium
iron garnet heterostructures indicates that interface-induced magnetic
exchange coupling (MEC) is accompanied by sizable Bychov-Rashba effect
\cite{Ferr_G_YIG_Wang15,AHE_G_Tang_18}. The breaking of inversion
symmetry in a honeycomb layer couples different SU(2) subspaces (spin
and sublattice) \cite{GRashba_09} and can drive the ferromagnetic
2D Dirac system through a topological phase transition to a Chern
insulator when the chemical potential is tuned inside the gap \cite{MEC_Gr_Theory_QAHE_Sun_10,Chen_11,QAHE_G_2}.
This system is predicted\textbf{ }to exhibit the quantum anomalous
Hall effect (QAHE), with transverse conductivity $\sigma_{\text{AH}}=2\,e^{2}/h$
\cite{MEC_Gr_Theory_QAHE_Sun_10}. However, much less is known about
the \emph{non-quantized regime} at finite carrier density. The latter
is the current experimental accessible regime \cite{Ferr_G_YIG_Wang15,AHE_G_Tang_18}.
Beyond the non-quantized part of the intrinsic contribution, the presence
of a Fermi surface makes the transverse (anomalous Hall) conductivity
depend nontrivially on spin-dependent scattering due to pseudospin-spin
coupling \cite{ferreira2014,Tuan16,milletari16,ASP_Huang_16}. Moreover,
in ultraclean heterostructures, the MEC and spin\textendash orbit
coupling (SOC) energy scales can easily reach the disorder-induced
broadening \cite{Zollner_PRB2016,Gmitra_PRB2016,Wang_NatComm2015},
thus questioning the use of standard approaches where SOC is treated
as a weak perturbation.

\begin{figure}
\begin{centering}
\includegraphics[width=1\columnwidth]{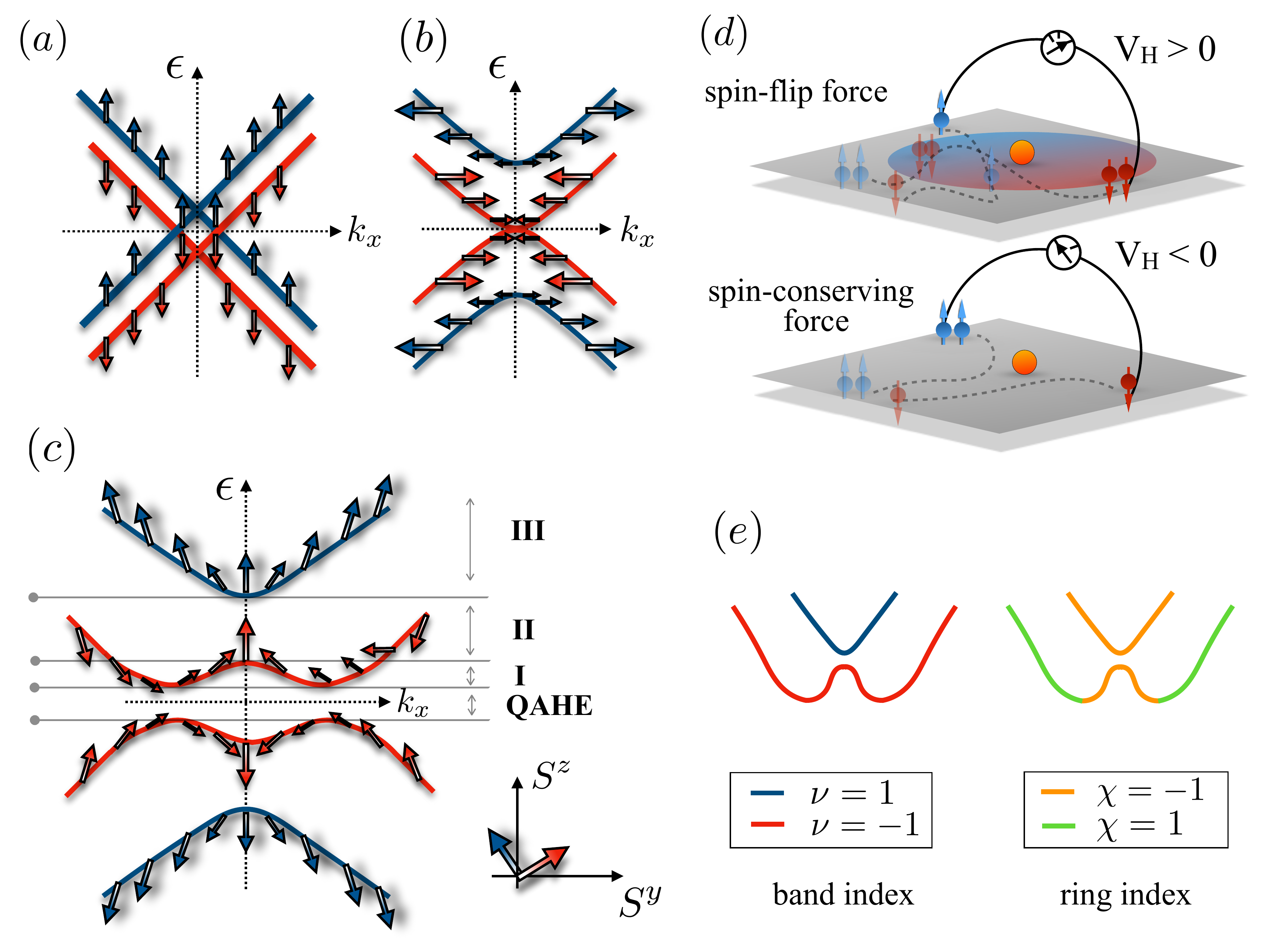}
\par\end{centering}
\centering{}\caption{\label{fig:energy_bands}(a-c) Energy bands and spin texture in systems
with (a) MEC (b) SOC and (c) MEC and SOC. For visualization purposes,
the bands are plotted along $\hat{k}_{x}$ (spins lie only in the
$yz$ plane). (d) Behavior of Hall conductivity due to competing spin-Lorentz
forces. The elastic scattering channel dominates in regime II and
III. (e) Band classification using band (ring) index $\nu=\pm1$ ($\chi=\pm1$).}
\end{figure}

In this Letter, we report an accurate theoretical study of charge
and spin transport in magnetized 2D Dirac systems by treating the
effects of strong MEC and SOC \emph{nonperturbatively} in the presence
of dilute random impurities. Our theory is valid for both weak (Born)
and strong (unitary) potential scattering and accounts for intervalley
processes from point defects. We find that the out-of-plane component
of the noncollinear spin texture $\mathbf{S}_{\mu\nu\mathbf{k}}$
{[}with $\mu(\nu)=\pm1$ pseudospin (spin chirality); Fig.~\ref{fig:energy_bands}{]}
activates a robust skew scattering mechanism, which determines the
behavior of leading Fermi surface contributions to the transverse
transport coefficients. The $\mathbf{k}$ modulation of the spin polarization
\textcolor{black}{manifests into a ubiquitous }\textcolor{black}{\emph{change
of sign }}\textcolor{black}{in the charge Hall conductivity as the
Fermi level approaches the majority spin band edge, which, as we argue
below, is a forerunner of the elusive QAHE }\cite{Ferr_G_YIG_Wang15,AHE_G_Tang_18,Leutenantsmeyer_16}\textcolor{black}{.}
Second, we predict that scattered electron waves with opposite polarization
(e.g., from within the 'Mexican hat' with \textcolor{black}{$S_{\mathbf{k}_{\pm}}^{z}\gtrless0$};
Fig.~\ref{fig:energy_bands}) have different transverse cross section
leading to net spin Hall current in the bulk \cite{SHE_RMP_2015}.
Such a spin Hall effect (SHE) in a 2D Dirac system with broken time
reversal symmetry can be seen as the reciprocal of the inverse spin
Hall effect discovered recently in ferromagnets\textcolor{blue}{{} \cite{Miao_PRL2013,Das_PRBR2017,Iihama_NatElet2018}}\textcolor{black}{.
The common stem of AHE and SHE implies the change of sign reveals
likewise in the spin Hall response, unveiling the possibility of }\textcolor{black}{\emph{reversing
the spin accumulation at the sample boundaries by gate voltage}}\textcolor{black}{.
The sign-change feature is found to be preserved when adding the Berry
curvature-dependent contributions to the AHE and SHE over a wide range
of parameters in samples with high mobility.}

\emph{Model}.\textemdash The low-energy Hamiltonian reads (we use
natural units $e\equiv1\equiv\hbar$, unless stated otherwise)
\begin{equation}
H=v\,\tau_{z}\,\boldsymbol{\Sigma}\cdot\mathbf{p}+\text{\ensuremath{\delta}}\,s_{z}+\lambda\,\tau_{z}\left(\Sigma_{x}s_{y}-\Sigma_{y}s_{x}\right)+V(\mathbf{x})\,,\label{eq:Hamiltonian}
\end{equation}
where $v$ is the Fermi velocity of massless Dirac fermions, $\delta$
($\lambda$) is the MEC (Bychov-Rashba) energy scale, and $V(\mathbf{x})$
is a disorder potential describing impurity scattering. Here, $\left\{ \boldsymbol{\tau},\boldsymbol{\Sigma},\mathbf{s}\right\} $
are Pauli matrices defined on valley, pseudospin and spin spaces,
respectively, and $\mathbf{p}=-i\boldsymbol{\nabla}$ is the 2D kinematic
momentum operator for states near the $K$ ($K^{\prime}$) point ($\tau_{z}=\pm1$).
This model describes magnetized graphene with $C_{6v}$ point group
symmetry \cite{Ferr_G_YIG_Wang15,AHE_G_Tang_18} and can be easily
extended to other ferromagnetic 2D materials, such as MoTe$_{2}$/EuO
 \cite{Qi_MoTe2_2015,Cheng_16,Yao_PRB2017}.\textbf{ }We consider
(nonmagnetic) matrix disorder with $V(\mathbf{x})=\sum_{i}\left(u_{0}1+u_{x}\tau_{x}\right)\delta(\mathbf{x}-\mathbf{x}_{i})$,
where $\{\mathbf{x}_{i}\}_{i=1...N}$ are random impurity positions
and $u_{0(x)}$ parameterizes the intravalley (intervalley) scattering
strength \cite{G_Review,ResonantScatt_G}. This choice allows us to
interpolate between ``smooth'' potentials in clean samples ($|u_{x}|\ll|u_{0}|$)
and the ``sharp defect'' limit of enhanced backscattering processes
($u_{x}\approx u_{0}$). The energy-momentum dispersion relation associated
to clean system $H_{0}=H-V(\text{\textbf{x}})$ reads 
\begin{equation}
\epsilon_{\mu\nu}(\mathbf{k})=\mu\sqrt{v^{2}k^{2}+M_{\nu}^{2}\left(k\right)}\,,\label{eq:dispersion}
\end{equation}
where $M_{\nu}(k)=\sqrt{2\lambda^{2}+\text{\ensuremath{\delta^{2}}}+2\nu\sqrt{\lambda^{4}+v^{2}k^{2}(\lambda^{2}+\text{\ensuremath{\delta^{2}}})}}$
is the SOC mass and $k=|\mathbf{k}|$ is the wavevector measured from
a Dirac point. Indices $\{\mu,\nu\}=\pm1$ define the carrier polarity
and the spin winding direction (Fig.\,\ref{fig:energy_bands}). 
\textcolor{black}{{} }In the absence of SOC, the Dirac cones are shifted
vertically, resulting in mixed electron\textendash hole states near
the Dirac point. For $\lambda\neq0,\,\text{\ensuremath{\delta}}=0$
(no MEC), the spectrum admits a spin-gap or \emph{pseudogap }region,
within which the spin and momentum of quasiparticles are locked at
right angles (Bychov-Rashba spin texture) \cite{GRashba_09,offidani17}.
The combination of SOC and MEC opens a gap and splits the Dirac spectrum
into 3 branches: \emph{regions I and III,} defined by $|\lambda\text{\ensuremath{\text{\ensuremath{\delta}}}|/\ensuremath{\sqrt{\lambda^{2}+\text{\ensuremath{\text{\ensuremath{\delta}}^{2}}}}}}\equiv\textrm{\ensuremath{\epsilon}}_{\textrm{I}}<|\epsilon|<\epsilon_{\textrm{II}}=|\text{\ensuremath{\delta}|}$
and $|\epsilon|>\sqrt{4\lambda^{2}+\text{\ensuremath{\text{\ensuremath{\delta}}^{2}}}}\equiv\epsilon_{\textrm{III}}$;
those energy regimes are characterized by a non-simply connected Fermi
surface allowing for scattering between states with different Fermi
momenta; and \emph{region II, }$\epsilon_{\textrm{II}}<|\epsilon|<\epsilon_{\textrm{III}}$,\emph{
}with only one band intersecting the Fermi level.\textcolor{black}{{}
For brevity, all functions are projected onto valley $\tau_{z}=1$
($K$ point).} The Bloch eigenstates read as 
\begin{equation}
|\psi_{\mu\nu\mathbf{k}}\left(\mathbf{x}\right)\rangle=\left(\begin{array}{c}
e_{\,}^{-\imath\phi_{\mathbf{k}}}\\
\imath\frac{\left(\epsilon_{\mu\nu}-\text{\ensuremath{\text{\ensuremath{\delta}}}}\right)^{2^{\,}}-v^{2}k^{2}}{2\,v\,k\,\lambda_{\,}}\\
\frac{\epsilon_{\mu\nu}-\text{\ensuremath{\text{\ensuremath{\delta}}^{\,}}}}{vk_{\,}}\\
\imath\frac{\left(\epsilon_{\mu\nu}-\text{\ensuremath{\text{\ensuremath{\delta}}}}\right)^{2^{\,}}-v^{2}k^{2}}{2\,\lambda\left(\epsilon_{\mu\nu}+\text{\ensuremath{\text{\ensuremath{\delta}}}}\right)}e^{\imath\phi_{\mathbf{k}}}
\end{array}\right)e^{\imath\mathbf{k}\cdot\mathbf{x}}\,,\label{eq:Eigenstates}
\end{equation}
where $\phi_{\mathbf{k}}$ is the wavevector polar angle. The noncoplanar
spin texture in momentum space highlights the competition between
different interactions: while the Bychov-Rashba effect favors in-plane
alignment, the exchange interaction tilts the spins out of the plane,
leading to a noncoplanar band polarization {[}Fig.\textcolor{teal}{~}\ref{fig:energy_bands}
(c){]}. The pronounced effects of symmetry breaking on the spin texture
has been highlighted in other systems, e.g. surface states of Bi thin
films \cite{Takayama_2011PRL}.\textbf{ }We underline here its impact
on relativistic transport: as shown below, the out-of-plane spin texture
$S_{\mu\nu\mathbf{k}}^{z}\equiv\frac{1}{2}\langle s_{z}\rangle_{\mu\nu\mathbf{k}}$
modulates  intrinsic and extrinsic transport contributions; even if
the electronic states are not fully spin polarized, it \textcolor{black}{will
prove useful to refer to }\textcolor{black}{\emph{effective}}\textcolor{black}{{}
spin-up ($S^{z}>0$) and spin-down ($S^{z}<0)$ states. We focus on
positive energies, $\epsilon>0$, and also $\lambda,\delta>0$, thus
fixing $\mu=1$ and omitting this index from the expressions. }

\emph{Spin texture-driven skew scattering.\textemdash{}} To assess
the dominant extrinsic transport contributions in the metallic regime
($\epsilon>\epsilon_{\textrm{I}}$), we solve the Boltzmann transport
equations (BTEs) for a spatially homogeneous system. The formalism
allows for the inclusion of a nonquantizing magnetic field and, more
importantly, for a transparent physical interpretation of the scattering
processes. For a controlled quantum diagrammatic treatment at the
$T$-matrix level, we refer the reader to the supplementary material
(SM) \cite{supplemMat}, where quantum side jump corrections are shown
to be subleading for typical (dilute) impurity concentrations. The
BTEs read as\begin{widetext}

\begin{equation}
\partial_{t}f_{\mathbf{k}_{\chi}}-e\left(\boldsymbol{\mathcal{E}}+\mathbf{v}\times\boldsymbol{\mathcal{B}}\right)\cdot\nabla_{\mathbf{k}}f_{\mathbf{k}_{\chi}}=2\pi n_{i}\sum_{\chi^{\prime}=\pm1}\int\frac{Sd^{2}\mathbf{k}^{\prime}}{(2\pi)^{2}}\left(f_{\mathbf{k}_{\chi^{\prime}}^{\prime}}\mathcal{T}_{\mathbf{k}_{\chi^{\prime}}^{\prime}\mathbf{k}_{\chi}}-f_{\mathbf{k}_{\chi}}\mathcal{T}_{\mathbf{k}_{\chi}\mathbf{k}_{\chi'}^{\prime}}\right)\,\delta(\epsilon_{\mathbf{k}_{\chi}}-\epsilon_{\mathbf{k}_{\chi^{\prime}}^{\prime}})\,,\label{eq:BTE}
\end{equation}
\end{widetext}where $f_{\mathbf{k}_{\chi}}=f_{\mathbf{k}_{\chi}}^{0}+\delta f_{\mathbf{k}_{\chi}}$
is the sum of the Fermi-Dirac distribution function and $\delta f_{\mathbf{k}_{\chi}}$,
the deviation from equilibrium. Moreover, $\boldsymbol{\mathcal{E}},\boldsymbol{\mathcal{B}}$
are external DC fields, $e$ is the elementary charge and $S$ is
the area. The right-hand side is the collision term describing\textcolor{red}{{}
}\textcolor{black}{single impurity scattering and }$n_{i}$ is the
impurity areal density.\textcolor{black}{{} }Subscripts $\chi,\chi^{\prime}=\pm1$
are \emph{ring}\textbf{\emph{ }}indices for the \emph{outer/inner
}Fermi surfaces associated with momenta $k_{\pm}=v^{-1}\{\epsilon^{2}+\text{\ensuremath{\delta^{2}}}\pm[\epsilon^{2}\lambda^{2}+(\epsilon^{2}-\lambda^{2})\ensuremath{\delta^{2}}]^{1/2}\}^{1/2}$;
Fig.\,\ref{fig:energy_bands}(d) \cite{CommentRingIndex}. Accounting
for possible scattering resonances due to the Dirac spectrum \cite{ferreira2014},
transition rates are evaluated by means of the $T$-matrix approach
i.e., $\mathcal{T}_{\mathbf{k}\mathbf{k}^{\prime}}=\left|\langle\mathbf{k}^{\prime}|t|\mathbf{k}\rangle\right|^{2}$,
where $t=\mathcal{V}/\left(1-g_{0}\,\mathcal{V}\right)$ with $\mathcal{V}=u_{0}+\tau_{x}u_{x}$
and $g_{0}=\int d^{2}\mathbf{k}/(4\pi^{2})\,(\epsilon-H_{0\mathbf{k}}+\imath\,0^{+})^{-1}$
is the integrated propagator. We start by considering $u_{x}=0$,
for which electrons undergo \emph{intra- }and \emph{inter-ring }scattering
processes in the same valley (see \cite{supplemMat} for a graphical
visualization).  Exploiting the Fermi surface isotropy, and momentarily
setting $\boldsymbol{\mathcal{B}}=0$, the exact solution to the linearized
BTEs ($\nabla_{\mathbf{k}}f_{\mathbf{k}_{\chi}}\rightarrow\nabla_{\mathbf{k}}f_{\mathbf{k}_{\chi}}^{0}$)
is 
\begin{align}
\delta f_{\mathbf{k}_{\chi}} & =-e\left(\frac{\partial f_{\mathbf{k}_{\chi}}^{0}}{\partial\epsilon}\right)\,\mathbf{v}_{\mathbf{k}_{\chi}}\cdot\left(\tau_{\chi}^{\parallel}\,\boldsymbol{\mathcal{E}}+\tau_{\chi}^{\perp}\,\hat{z}\times\boldsymbol{\mathcal{E}}\right)\,,\label{eq:ansatz}
\end{align}
with $\mathbf{v}_{\mathbf{k}_{\chi}}=\nabla_{\mathbf{k}}\epsilon_{\chi}(\mathbf{k})$.
In the above, $\tau_{\chi}^{\varsigma}=\tau_{\chi}^{\varsigma}(\epsilon,\lambda,\text{\ensuremath{\delta}},u_{0},n_{i})$
are the longitudinal ($\varsigma=\parallel$) and transverse ($\varsigma=\perp$)
transport times given by
\begin{align}
\boldsymbol{\tau}^{\parallel} & =-2\left(\hat{\Lambda}+\hat{\Upsilon}\hat{\Lambda}^{-1}\hat{\Upsilon}\right)^{-1}\boldsymbol{1}\,,\,\,\,\boldsymbol{\tau}^{\perp}=\hat{\Lambda}^{-1}\hat{\Upsilon}\,\boldsymbol{\tau}^{\parallel}\,,\label{eq:tauPar}
\end{align}
where $\boldsymbol{\tau}^{\varsigma}=(\tau_{\chi}^{\varsigma},\tau_{\bar{\chi}}^{\varsigma})^{\text{t}}$,
$\hat{\Lambda}=((\Lambda_{\chi}^{-},\Lambda_{\chi}^{+})^{\text{t}},(\Lambda_{\bar{\chi}}^{-},-\Lambda_{\bar{\chi}}^{+})^{\text{t}})$,
$\boldsymbol{1}=(1,0)^{\text{t}}$ and $\bar{\chi}\equiv-\chi$ ($\hat{\Upsilon}$
is obtained from $\hat{\Lambda}$ via the substitution $\Lambda_{\chi}^{\pm}\to\Upsilon_{\chi}^{\mp}$).
The kernel\textcolor{black}{s $\Lambda_{\chi}^{\pm}$ }and $\Upsilon_{\chi}^{\pm}$\textcolor{black}{{}
are cumbersome functions of symmetric and skew cross sections defined
by }$\boldsymbol{\Gamma}_{\chi\chi^{\prime}}=\frac{n_{i}}{2\pi}\int Sd^{2}\text{\textbf{k}}^{\prime}\,\mathcal{T}_{\mathbf{k}_{\chi^{\prime}}^{\prime}\mathbf{k}_{\chi}}\,\left\{ 1,\cos\phi,\sin\phi\right\} ^{\textrm{t}}$
with $\phi=\phi_{\text{\textbf{k}}^{\prime}}-\phi_{\text{\textbf{k}}}$
\cite{supplemMat}. Considering the two valleys, the general solution
involves 16 cross sections. The exact form of the kernels is essential
to correctly determine the energy dependence of the conductivity tensor.
As shown in SM \cite{supplemMat}, including a magnetic field $\boldsymbol{\mathcal{B}}=\mathcal{B}\,\hat{z}$
only requires the substitution $\Gamma_{\chi\chi}^{\text{sin}}\to\Gamma_{\chi\chi}^{\text{sin}}+\omega_{\chi}$,
where $\omega_{\chi}=k_{\chi}v_{\chi}^{-1}\mathcal{B}$ is the cyclotron
frequency associated with the ring states. At $T=0$, accounting for
the valley degeneracy, we obtain the transverse response functions 

\begin{equation}
\sigma_{\perp}^{c,s}(\mathcal{B},\epsilon)=\frac{-e}{h}\sum_{\chi=\pm1}k_{\chi}(\epsilon)\,\langle J_{c,s}(\epsilon)\rangle_{\chi}\,\tau_{\chi}^{\perp}(\epsilon,\mathcal{B})\,,\label{eq:solCond}
\end{equation}
where $\langle J_{c,s}(\epsilon)\rangle_{\chi}=-e\,\langle\{1,s_{z}/2\}\,\boldsymbol{\Sigma}\cdot|\hat{z}\times\boldsymbol{\hat{\mathcal{E}}}|\rangle_{\chi}$
denotes the equilibrium transverse charge (spin) current of planewave
states in the $\chi$ ring. The skew cross sections (and hence $\tau_{\chi}^{\perp}$)
are found to be nonzero (except for isolated points) and thus, in
the dilute regime, one has $\sigma_{\perp}^{c,s}\propto n_{i}^{-1}$,
which is a signature of skew scattering \cite{milletari16}.\textbf{
}As discussed below, the energy dependence of the skew cross sections
is very marked, reflecting the out-of-plane  spin texture of conducting
electrons. For simplicity, in what follows, we work at saturation
field $\mathcal{B}\ge\mathcal{B}_{\text{sat}}$ such that the transverse
responses coincide with their ``anomalous'' parts, that is, $\left.\sigma_{\text{AH(SH)}}(\mathcal{B}_{\text{sat}},\epsilon)\right|_{\delta}\simeq\left.\sigma_{\perp}(0,\epsilon)\right|_{\delta_{\textrm{sat}}}$,
where $\delta_{\textrm{sat}}=\delta(M_{z}(\mathcal{B}_{\text{sat}}))$
and $M_{z}$ is the magnetization.

\begin{figure}
\begin{singlespace}
\includegraphics[width=1\columnwidth]{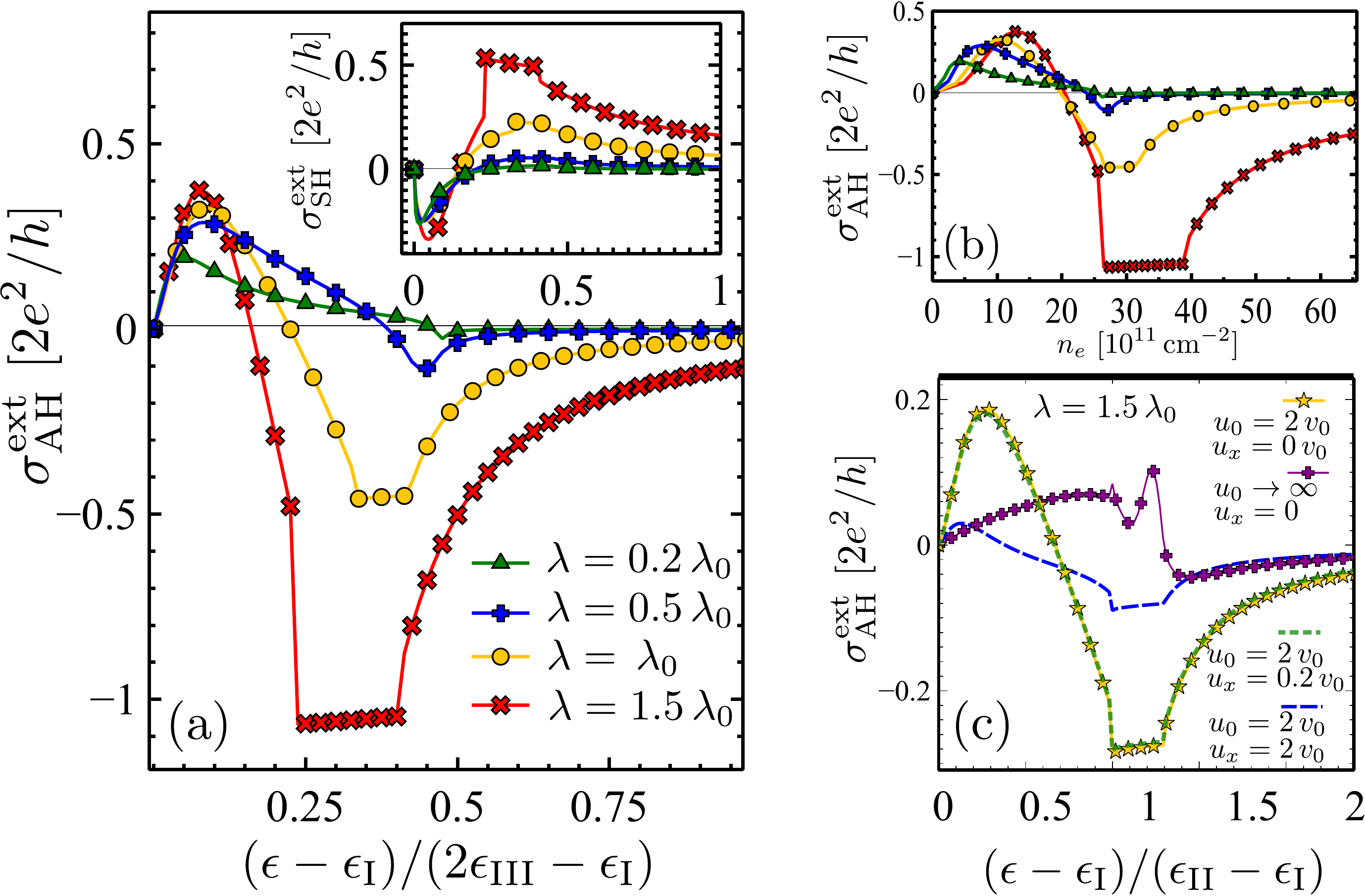}
\end{singlespace}

\caption{\label{fig:-as-a}Energy dependence of AHE and SHE. (a) $\sigma_{\perp}^{c,s}(\epsilon)$
is the result of the competing effective spin Lorentz forces as discussed
in the main text {[}see also Fig.\,\eqref{fig:energy_bands}(e){]}.
The \emph{change of sign} is more prominent for large SOC. (b) $\sigma_{\perp}^{c}(n_{e})$
where $n_{e}=\pi\{k_{+}^{2}-k_{-}^{2},\,k_{+}^{2},\,k_{+}^{2}+k_{-}^{2}\}$,
respectively, in regions I, II, and III. (c) While less evident in
the unitary limit, the change of sign is robust across all scattering
regimes. $\delta=30$ meV, $\lambda_{0}=\delta/3$, $v_{0}=u_{0}=1\,\textrm{eV}\cdot\textrm{nm}^{2},\,u_{x}=0$
and $n_{0}=10^{12}\text{cm}^{-2}$. }
\end{figure}
 \emph{The change of sign.\textemdash }Focusing on the regime $\lambda\lesssim\delta$,
we show how, approaching low carrier density, electrons undergoing
\emph{spin-conserving}  and \emph{spin-flip} scattering processes
determine a change of sign in $\sigma_{\perp}^{c,s}$. For the sake
of illustration, we assume weak scatterers $|g_{0}u_{0}|\ll1$ and
restrict the analysis to intraring transitions within the outer ring:
$\mathbf{k}_{+}\to\mathbf{k}_{+}^{\prime}$ (see additional discussions
\cite{supplemMat}). A first scenario for the change of sign is as
follows. First, we note that as $k$ is increased from $k=0$, electron
states in the lower band $\nu=-1$ progressively change their spin
orientation from effective spin-up to -down states (see Fig.\,\ref{fig:energy_bands}).
Starting from $\epsilon_{\text{I}}$, varying $\epsilon$ instead,
it can be verified that the same occurs within the outer ring, such
that by tuning $\epsilon$ one can switch between states with opposite
spin polarization. As effective up/down states are associated with
an opposite effective spin Lorentz force (i.e., skew cross sections
with opposite signs), this also means conducting electrons can be
selectively deflected towards opposite boundaries of the system. The
associated anomalous Hall (AH) voltage and SHE spin accumulation will
then display the characteristic change of sign {[}Fig.\,\ref{fig:-as-a}(a){]}. 

A second scenario involves the spin-flip force and does not require
changing the polarization of carriers. Instead, what changes when
varying $\epsilon$ is the ratio of spin-flip to elastic skew cross
sections. This also produces a change of sign as depicted in Fig.\,\ref{fig:energy_bands}(e);
the fate of the transverse conductivity will depend ultimately on
the competition between the two effective spin Lorentz forces (see
SM \cite{supplemMat}).\textbf{ }The change of sign in $\sigma_{\perp}^{c,s}$
is a persistent feature as long as SOC and MEC are comparable {[}Fig.\,\ref{fig:-as-a}(a){]}.
In that case, the noncollinear spin texture is well developed, such
that, on one hand, it is possible to interchange between effective
spin-up and -down states $"S"="\uparrow,\downarrow"=-"\bar{S}"$ using
a gate voltage, and, on the other, both spin-conserving $\langle"S"|V(\text{\textbf{x}})|"S"\rangle$
and spin-flip $\langle"\bar{S}"|V(\text{\textbf{x}})|"S"\rangle$
scattering matrix elements are non-zero. Asymptotically, $\epsilon\gg\delta,\lambda$,
the AH signal must vanish due to the opposite spin orientation of
electron states belonging to $\nu=\pm1$ bands, which produce vanishing
small total magnetization $S^{z}(\epsilon)=\sum_{\chi}S_{\chi}^{z}\ll1$.
In comparison, the staggered field experienced by charge carriers
$S_{\textrm{stag}}^{z}(\epsilon)=\sum_{\chi}\chi S_{\chi}^{z}$ has
slower asymptotic decay (for $\lambda\lesssim\delta$), implying that
the SHE is more robust than the AHE. 

\emph{Approaching the QAHE.\textemdash }The system realizes the QAHE
provided the gap remains robust against disorder, $\sigma_{\perp}^{c}(\epsilon<\epsilon_{\text{I}})=2e^{2}/h,\,\,\sigma_{\perp}^{s}(\epsilon<\epsilon_{\text{I}})=0$.
In the metallic regime, the Berry curvature of occupied states also
provide a (nonquantized) intrinsic contribution to the transverse
conductivity. Below, we discuss how robust is the change of sign to
the inclusion of other factors and also how the quantized region is
approached. First, consider that in the strong scattering limit, $|g_{0}u_{0}|\gg1$,
the rate of inter-ring transitions increases and the one-ring scenario
presented above might break down. However, as shown in Fig.\,\ref{fig:-as-a}(c)
the change of sign is still visible. In real samples, structural defects
and short-range impurities, such as hydrocarbons \cite{Resonant_Scatt_Ni_10},
induce scattering between inequivalent valleys, thereby opening the
backscattering channel \cite{G_Review}. In fact, spin precession
measurements in graphene with interface-induced SOC indicate that
the in-plane spin dynamics is sensitive to intervalley scattering
\cite{Benitez_17,Cummings_17,Ghiasi_17}. To determine the impact
of intervalley processes on DC transport, we solved the BTEs for arbitrary
ratio $u_{x}/u_{0}$. Figure\,\ref{fig:-as-a}(c) shows the AH conductivity
for selected values of $u_{x}$ (dashed lines). $\sigma_{\perp}^{c}$
is strongly impacted showing a 50\% reduction when intra- and intervalley
scattering processes are equally probable ($u_{x}=u_{0}$). However,
the sign change in $\sigma_{\perp}$, approaching the majority spin
band edge $\epsilon\approx\epsilon_{\textrm{II}}$ is still clearly
visible. Further analysis are given in SM \cite{supplemMat}, where
we also analyze the impact of thermal fluctuations, concluding that
the features described above are persistent up to $k_{\text{B}}T\approx k_{\text{B}}T_{\text{room}}/2\simeq12\,\text{meV}$
for $\lambda\approx\delta=30$~meV. A thorough numerical analysis
in the strong SOC regime provides an estimation for $\epsilon_{0}$
defined as $\sigma_{\perp}(\epsilon_{0})=0$,
\begin{equation}
\epsilon_{0}=a\,\epsilon_{\text{I}}+b\,\epsilon_{\text{III}}\,,\label{eq:EstimationE}
\end{equation}
with $a\simeq0.3$\textendash 0.4 and $b\simeq0.6$\textendash 0.8.
This relation shows that the knowledge of $\text{\ensuremath{\delta}}$
(e.g., from the Curie temperature \cite{Ferr_G_YIG_Wang15}) allows
to estimate the SOC strength directly from the gate voltage dependence
of the AH resistance. The values $\sigma_{\perp}\approx0.1-1$ ($e^{2}/h$)
are compatible with the measurements in Refs.\,\cite{Ferr_G_YIG_Wang15,AHE_G_Tang_18},
for a reasonable choice of parameters, $0\le\lambda,\delta\le30\,\text{meV}$
in the dirty regime with $n_{i}=10^{12}\,\text{cm}^{2}$ and $u_{0}\sim(0.1,1)\,\text{eV}\cdot\textrm{nm}^{2}$.
In high mobility samples, our theory predicts that the robust skew
scattering contribution with $\sigma_{\perp}\propto n_{i}^{-1}$ results
in much larger values $\sigma_{\perp}\approx10-100$ ($e^{2}/h$).

\begin{figure}
\includegraphics[width=1\columnwidth]{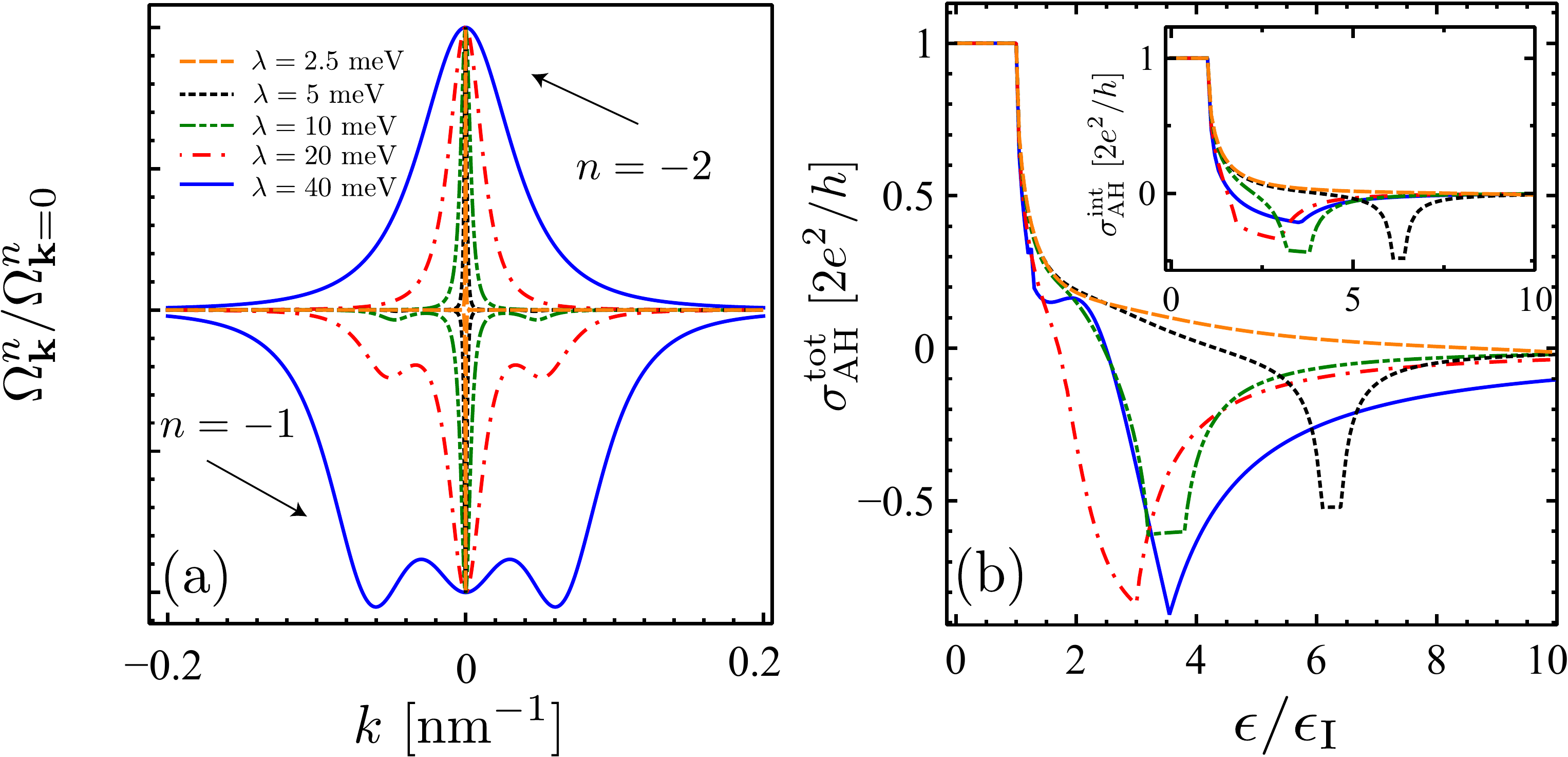}

\caption{\label{fig:IMC-and-competition} Intrinsic contribution and total
AH conductivity at selected SOC values. (a) The Berry curvatures of
hole bands $n=-1,-2$. Note that $\Omega_{n=-1}(\mathbf{k})$ develops
additional ``hot spots'' as SOC is increased. (b) The total $\sigma_{\perp}^{\text{tot}}=\sigma_{\perp}^{\text{}}+\sigma_{\perp}^{\text{int}}$;
same legend as in (a). (b) Adding the intrinsic contribution (inset)
to $\sigma_{\perp}$ leaves the estimate for $\epsilon_{0}$ virtually
unaffected {[}Eq.\,\eqref{eq:EstimationE}{]}. Parameters: $u_{0}=10\text{ \textrm{eV}\ensuremath{\cdot\textrm{nm}^{2}}},$
$n_{i}=10^{12}\text{ cm}^{-2}$ and $\delta=30\text{ meV}$. }
 
\end{figure}

\emph{Intrinsic contribution and total AH conductivity.\textemdash }We
now report our results for the intrinsic contribution. Previous studies\textemdash where
the topological nature of the model was also firstly pointed out \cite{MEC_Gr_Theory_QAHE_Sun_10}\textemdash tackled
the problem numerically, also with a focus in the regime $\text{\ensuremath{\delta}}\gg\lambda$.
We go beyond this limitation performing an exact analytic evaluation
of the intrinsic AH conductivity. Starting from the chiral eigenstates
of Eq.\,\eqref{eq:Eigenstates}, we obtain the Berry curvature of
the bands as $\Omega_{\mathbf{k}}^{n}=\left(\nabla_{\mathbf{k}}\times\boldsymbol{\mathcal{A}}_{\mathbf{k}}^{n}\right)_{z}$,
where $\boldsymbol{\mathcal{A}}_{\mathbf{k}}^{n}=-i\langle n\mathbf{k}|\nabla_{\mathbf{k}}|n\mathbf{k}\rangle$
and $n\equiv(\mu,\nu)$ is a combined band index \cite{CommentIndex}.
The transverse conductivity is obtained via integration of the Berry
curvatures \cite{TKNNFormula}, $\sigma_{\perp}^{\textrm{int}}=\sum_{n}\sum_{\mathbf{k}}\Omega_{\mathbf{k}}^{n}\,f_{k_{n}}^{0}$.
Note that $\sum_{n}\Omega_{\mathbf{k}}^{n}=0$ and $\sum_{\mathbf{k}}\sum_{n<0}\Omega_{\mathbf{k}}^{n}=2e^{2}/h$,
which is the case when $\epsilon$ is tuned into the gap. The full
form of $\boldsymbol{\mathcal{A}}_{\mathbf{k}}^{n}$ is reported in
Ref.\,\cite{supplemMat}, where we also show that the intrinsic contribution
can be equivalently obtained from the clean limit of the Kubo\textendash Streda
formula. The result is plotted in the inset of Fig.\,\ref{fig:IMC-and-competition}(b),
while Fig.\,\ref{fig:IMC-and-competition}(a) shows the opposite-in-sign
Berry curvatures for the bands $n=-1,-2$. Similarly to the situation
presented for the extrinsic contribution, we find the intrinsic term
also presents a peculiar change of sign \textcolor{black}{under the
same condition $\lambda>\lambda_{c}\approx\delta/6$ {[}}see Fig.~\ref{fig:-as-a}
(a){]}, where $\lambda_{c}$ is a critical value for the Bychov-Rashba
strength. The effect in this case is ascribed to the profile of the
Berry curvatures; in particular, in the electron sector the change
of sign happens for $\epsilon=\tilde{\epsilon}_{0}$ solution of the
self-consistent equation
\begin{equation}
\left.I_{1}\right|_{k_{-}}^{k_{+}}\left(\tilde{\epsilon}_{0}\right)+\theta_{\tilde{\epsilon}_{0},\epsilon_{\textrm{II}}}\left.I_{1}\right|_{0}^{k_{+}}\left(\tilde{\epsilon}_{0}\right)+\theta_{\tilde{\epsilon}_{0},\epsilon_{\textrm{III}}}\left.I_{2}\right|_{0}^{k_{-}}\left(\tilde{\epsilon}_{0}\right)=-\frac{2e^{2}}{h}\,,
\end{equation}
where $\left.I_{i}\right|_{a}^{b}\left(\epsilon\right)\equiv\int_{a(\epsilon)}^{b(\epsilon)}dk\,k\,\Omega_{\mathbf{k}}^{i}$
and $\theta_{\epsilon_{a},\epsilon_{b}}=\theta\left(\epsilon_{a}-\epsilon_{b}\right)$
is the Heaviside step function. In Fig.\,\ref{fig:IMC-and-competition}(b)
we show the total AH conductivity, given by $\sigma_{\perp}^{\text{tot}}=\sigma_{\perp}^{\textrm{int}}+\sigma_{\perp}^{c}$.
Remarkably, we find $\tilde{\epsilon}_{0}\simeq\epsilon_{0}$, such
that our estimate for the AHE reversal energy ($\epsilon_{0}$) in
Eq.\,\eqref{eq:EstimationE} is still accurate when adding all contributions
(cf. Fig.\,\ref{fig:IMC-and-competition}(b) and inset).\textcolor{teal}{{}
}This robust energy dependence in the AHE/SHE transverse response
functions connects the skew scattering mechanism, unveiled in this
work, to the intrinsic properties of magnetized 2D Dirac bands.

\emph{Acknowledgments.\textemdash }The authors are grateful to Chunli
Huang, Denis Kochan and Mirco Milletarì for useful discussions. We
thank Stuart A. Cavill and Roberto Raimondi for critically reading
the manuscript and for helpful comments. A.F. gratefully acknowledges
the financial support from the Royal Society (U.K.) through a Royal
Society University Research Fellowship. M.O. and A.F. acknowledge
funding from EPSRC (Grant Ref: EP/N004817/1).

\newpage{}

\newpage{}

\onecolumngrid
\newpage{}

\section*{SUPPLEMENTARY INFORMATION}

In this Supplementary Information, we derive the exact solution of
linearized BTEs for 2D magnetic Dirac bands presented in the Letter
and discuss a number of additional results, including a calculation
of quantum side-jump correction and the analytical form of the Berry
curvature in the full 4-band model. The equivalence between the extrinsic
contribution obtained from Boltzmann transport equations and the Kubo\textendash Streda
formalism is also established.

\tableofcontents{}

\section{SEMICLASSICAL THEORY}

\subsection{LINEARIZED BOLTZMANN TRANSPORT EQUATIONS }

In what follows, we derive the analytical form of the nonequilibrium
distribution function for intravalley scattering potentials. For brevity,
we work at fixed Fermi energy, $\epsilon>0$. The scattering probability
is given by
\begin{equation}
W_{\mathbf{k}_{\chi},\mathbf{k}_{\chi^{\prime}}^{\prime}}=2\pi n_{i}\,\mathcal{T}_{\mathbf{k}_{\chi}\mathbf{k}_{\chi^{\prime}}^{\prime}}\,\delta(\epsilon_{\text{\textbf{k}}_{\chi}}-\epsilon_{\text{\textbf{k}}_{\chi^{\prime}}^{\prime}})=2\pi n_{i}\left|\langle\mathbf{k}_{\chi^{\prime}}^{\prime}|t|\mathbf{k}_{\chi}\rangle\right|^{2}\delta(\epsilon_{\text{\textbf{k}}_{\chi}}-\epsilon_{\text{\textbf{k}}_{\chi^{\prime}}^{\prime}})\,,
\end{equation}
where all the quantities appearing in the last equation are defined
in the main text. Throughout this supplemental material, we also employ
the following definitions
\begin{equation}
(d\mathbf{k})=d^{2}\mathbf{k}/4\pi^{2}\,,\quad\gamma_{0}\equiv\mathbb{I}_{8\times8}=\tau_{0}\Sigma_{0}s_{0},\quad\gamma_{\text{KM}}=\tau_{0}\Sigma_{z}s_{z}\quad\textrm{and}\quad\gamma_{r}=\tau_{z}\left(\boldsymbol{\Sigma}\times\boldsymbol{s}\right)\cdot\hat{z},\label{eq:def1}
\end{equation}
The wavefunctions are expressed in the basis:
\begin{equation}
(A^{\uparrow}K,\,A^{\downarrow}K,\,B^{\uparrow}K,\,B^{\downarrow}K,\,B^{\uparrow}K^{\prime},\,B^{\downarrow}K^{\prime},\,A^{\uparrow}K^{\prime},\,A^{\downarrow}K^{\prime})^{\text{t}}\,.\label{eq:Basis}
\end{equation}

\subsubsection{Exact solution in zero magnetic field}

\begin{figure}
\includegraphics[width=0.9\columnwidth]{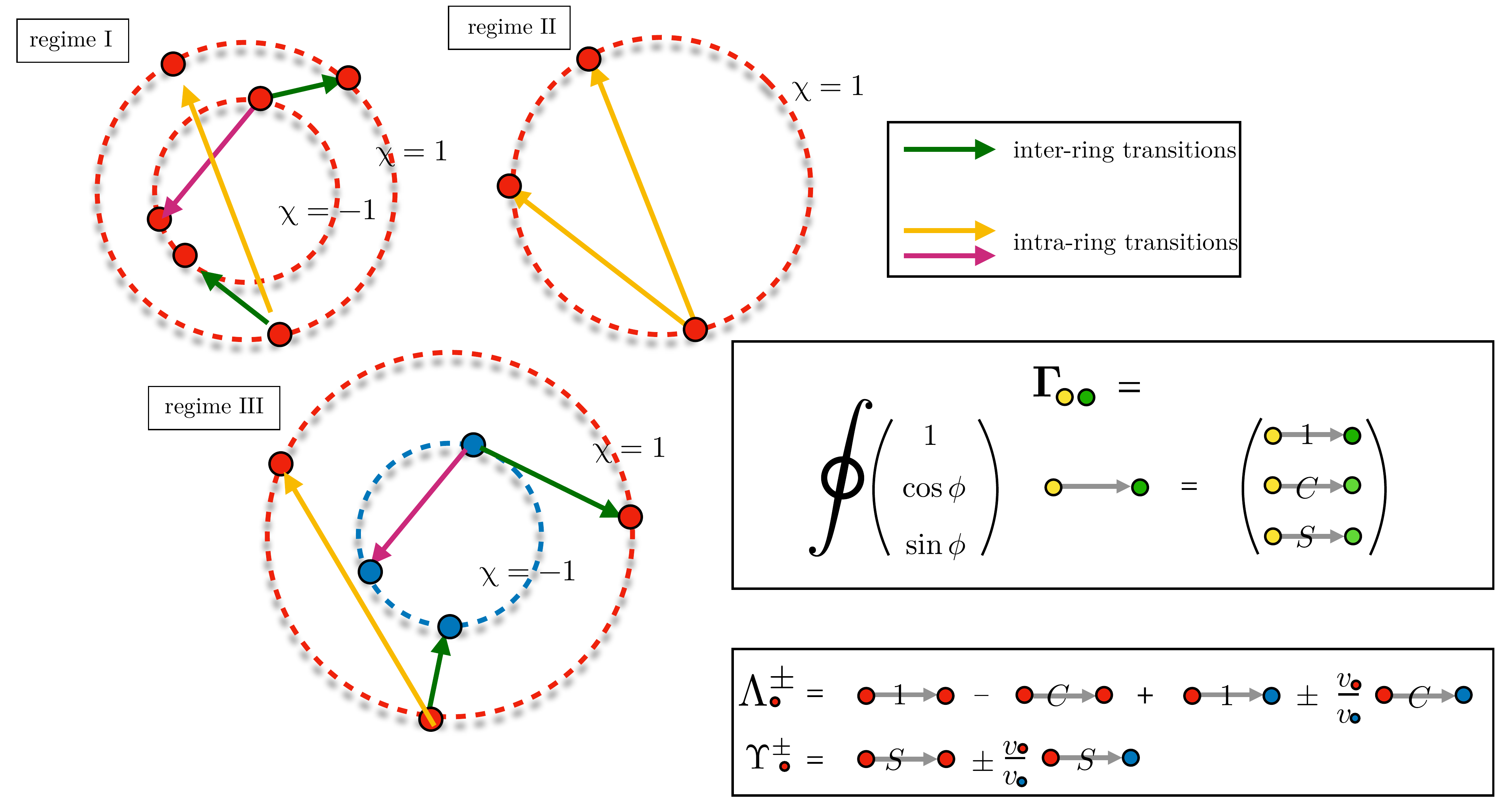}

\caption{\label{fig:Graphic-visualization-of}Graphic visualization of the
different impurity scattering processes in the 2D Dirac model in different
energy $\epsilon$ regimes (I, II and III; see Fig. 1; main text).
We report graphically the different relaxation rates $\Lambda_{\chi}^{\pm},\Upsilon_{\chi}^{\pm}$
mentioned in the main text. Colored dots are to be identified with
the indices $\chi$. When a generic scattering amplitude (grey segment
connecting yellow and green dots) is integrated over the angle, it
gives rise to different components of $\boldsymbol{\Gamma}_{\chi\chi^{\prime}}$
depending to which trigonometric function it is contracted with. Combinations
of the various components yield the different relaxation rates. }
\end{figure}

Without loss of generality, we take the electric field oriented along
the $\hat{x}$ direction. In the steady state of the linear response
regime, the left-hand side of the Boltzmann transport equations (BTEs)
{[}Eq.\,(4); main text{]} reads as \cite{Comment1}
\begin{equation}
-\boldsymbol{\mathcal{E}}\cdot\nabla_{\text{\textbf{k}}}f_{\text{\textbf{k}}_{\chi}}^{0}=-\boldsymbol{\mathcal{E}}\cdot\text{\textbf{v}}_{\text{\textbf{k}}_{\chi}}\left(\frac{\partial f_{\text{\textbf{k}}_{\chi}}^{0}}{\partial\epsilon}\right)=-\varsigma_{v}\left|\boldsymbol{\mathcal{E}}\right|v_{k_{\chi}}\cos\phi_{\text{\textbf{k}}}\left(\frac{\partial f_{\text{\textbf{k}}_{\chi}}^{0}}{\partial\epsilon}\right)\,,
\end{equation}
where $v_{k_{\chi}}$ is the band velocity of the $\chi$ Fermi ring
and $\varsigma_{v}$ is its sign, $\cos\phi_{\text{\textbf{k}}}=k_{x}/k$
and $f_{\text{\textbf{k}}_{\chi}}^{0}=(1+\text{Exp}[(\epsilon_{\text{\textbf{k}}_{\chi}}-\epsilon)/k_{B}T])^{-1}$.
To solve the BTEs, we make use of the ansatz Eq.\,(5) of the main
text, 

\begin{align}
\delta f_{\text{\textbf{k}}\chi} & =-\left(\frac{\partial f_{\text{\textbf{k}}_{\chi}}^{0}}{\partial\epsilon}\right)\left(\tau_{\chi}^{\parallel}\boldsymbol{\mathcal{E}}+\tau_{\chi}^{\perp}\,\hat{z}\times\boldsymbol{\mathcal{E}}\right)\cdot\text{\textbf{v}}_{\text{\textbf{k}}_{\chi}}\label{eq:ansatz-1}\\
 & =-\varsigma_{v}\,\mathcal{E}\left(\frac{\partial f_{\text{\textbf{k}}_{\chi}}^{0}}{\partial\epsilon}\right)v_{k_{\chi}}\left(\tau_{\chi}^{\parallel}\cos\phi_{\mathbf{k}_{\chi}}+\tau_{\chi}^{\perp}\,\sin\phi_{\mathbf{k}_{\chi}}\right)\,.
\end{align}
In regime II, only intra-ring processes are allowed, whereas in regime
I and III, one needs to take into account inter-ring transitions (Fig.
1; main text). For fixed index $\chi$, we separate intra-ring $(\chi\chi)$
and inter-ring $(\chi\bar{\chi})$ processes
\begin{align}
S\left[f_{\text{\textbf{k}}_{\chi}}\right] & =S^{\text{intra}}\left[f_{\text{\textbf{k}}_{\chi}}\right]+S^{\text{inter}}\left[f_{\text{\textbf{k}}_{\chi}}\right]\,,\\
S^{\text{intra}}\left[f_{\text{\textbf{k}}_{\chi}}\right] & =-\int(d^{2}\text{\textbf{k}}^{\prime})\left(f_{\text{\textbf{k}}_{\chi}}W_{\text{\textbf{k}}_{\chi},\text{\textbf{k}}_{\chi}^{\prime}}-f_{\text{\textbf{k}}_{\chi}^{\prime}}W_{\text{\textbf{k}}_{\chi}^{\prime},\text{\textbf{k}}_{\chi}}\right)\,,\\
S^{\text{inter}}\left[f_{\text{\textbf{k}}_{\chi}}\right] & =-\int(d^{2}\text{\textbf{k}}^{\prime})\left(f_{\text{\textbf{k}}_{\chi}}W_{\text{\textbf{k}}_{\chi},\text{\textbf{k}}_{\bar{\chi}}^{\prime}}-f_{\text{\textbf{k}}_{\bar{\chi}}^{\prime}}W_{\text{\textbf{k}}_{\bar{\chi}},\text{\textbf{k}}_{\chi}^{\prime}}\right)\,.
\end{align}
The different scattering probabilities are
\begin{align}
W_{\text{\textbf{k}}_{\chi},\text{\textbf{k}}_{\chi}^{\prime}} & =2\pi n_{i}\,\delta(\epsilon_{\text{\textbf{k}}_{\chi}-}\epsilon_{\text{\textbf{k}}_{\chi}^{\prime}})\,\mathcal{T}_{\chi\chi}(\phi)\,,\\
W_{\text{\textbf{k}}_{\chi},\text{\textbf{k}}_{\bar{\chi}}^{\prime}} & =2\pi n_{i}\,\delta(\epsilon_{\text{\textbf{k}}_{\chi}-}\epsilon_{\mathbf{k^{\prime}}_{\bar{\chi}}})\,\mathcal{T}_{\chi\bar{\chi}}(\phi)\,,\\
W_{\text{\textbf{k}}_{\chi}^{\prime},\text{\textbf{k}}_{\chi}} & =2\pi n_{i}\,\delta(\epsilon_{\text{\textbf{k}}_{\chi}^{\prime}-}\epsilon_{\mathbf{k}_{\chi}})\,\mathcal{T}_{\chi\chi}(-\phi)\,,\\
W_{\text{\textbf{k}}_{\bar{\chi}}^{\prime},\text{\textbf{k}}_{\chi}} & =2\pi n_{i}\,\delta(\epsilon_{\text{\textbf{k}}_{\bar{\chi}}^{\prime}-}\epsilon_{\mathbf{k}_{\chi}})\,\mathcal{T}_{\bar{\chi}\chi}(-\phi)\,.
\end{align}
It will be useful in the following to work with the symmetric and
antisymmetric components
\begin{equation}
\mathcal{T}_{\chi\chi^{\prime}}\left(\pm\phi\right)=\mathcal{T}_{\chi\chi^{\prime}}^{s}\pm\mathcal{T}_{\chi\chi^{\prime}}^{a}\,,
\end{equation}
along with the trigonometric relations
\begin{align}
\cos\left(\alpha+\beta\right) & =\cos\alpha\cos\beta-\sin\alpha\sin\beta\,,\\
\sin\left(\alpha+\beta\right) & =\sin\alpha\cos\beta+\sin\beta\cos\alpha\,,
\end{align}
to recast Eq.\,\eqref{eq:ansatz-1} into the form
\begin{align}
f_{\text{\textbf{k}}_{\chi}^{\prime}} & =-\varsigma_{v}\,\mathcal{E}\,v_{k_{\chi}}\frac{\partial f_{\text{\textbf{k}}_{\chi}^{\prime}}^{0}}{\partial\epsilon}\left[\cos\phi_{\text{\textbf{k}}_{\chi}}\left(\tau_{\chi}^{\parallel}\cos\phi+\tau_{\chi}^{\perp}\sin\phi\right)+\sin\phi_{\text{\textbf{k}}_{\chi}}\left(\tau_{\chi}^{\perp}\cos\phi-\tau_{\chi}^{\parallel}\sin\phi\right)\right]\,.
\end{align}
The intra-ring integrals now reduce to
\begin{align}
I_{1}^{\text{intra}} & \equiv\int(d^{2}\text{\textbf{k}}^{\prime})f_{\text{\textbf{k}}_{\chi}}W_{\text{\textbf{k}}_{\chi},\text{\textbf{k}}_{\chi}^{\prime}}=2\pi n_{i}\,f_{\text{\textbf{k}}_{\chi}}N(\epsilon_{\text{\textbf{k}}_{\chi}})\int\frac{d\phi}{2\pi}\mathcal{T}_{\chi\chi}\left(\phi\right)\,,\\
\nonumber \\
I_{2}^{\text{intra}} & \equiv\int(d^{2}\text{\textbf{k}}^{\prime})f_{\text{\textbf{k}}_{\chi}^{\prime}}W_{\text{\textbf{k}}_{\chi}^{\prime},\text{\textbf{k}}_{\chi}}\\
 & =2\pi n_{i}\,F_{\chi}\,v_{k_{\chi}}\,N(\epsilon_{\text{\textbf{k}}_{\chi}})\left\{ c_{1,\boldsymbol{\text{\textbf{k}}_{\chi}}}\int\frac{d\phi}{2\pi}\cos\phi\,\mathcal{T}_{\chi\chi}\left(-\phi\right)+c_{2,\text{\textbf{k}}_{\chi}}\int\frac{d\phi}{2\pi}\sin\phi\,\mathcal{T}_{\chi\chi}\left(-\phi\right)\right\} \,,
\end{align}
where $N(\epsilon_{\mathbf{k}})=\frac{1}{2\pi}\frac{k}{v_{k}}$ is
the density of states and
\begin{equation}
\{c_{1,\mathbf{\text{\textbf{k}}_{\chi}}},\,c_{2,\text{\textbf{k}}_{\chi}}\}=\{\cos\phi_{\text{\textbf{k}}_{\chi}}\tau_{\chi}^{\parallel}+\sin\phi_{\text{\textbf{k}}_{\chi}}\tau_{\chi}^{\perp},\,\cos\phi_{\text{\textbf{k}}_{\chi}}\tau_{\chi}^{\perp}-\sin\phi_{\text{\textbf{k}}_{\chi}}\tau_{\chi}^{\parallel}\}\,,\quad\textrm{and}\quad F_{\chi}=-\varsigma_{v}\,\mathcal{E}\,\frac{\partial f_{\text{\textbf{k}}_{\chi}}^{0}}{\partial\epsilon}\,.\label{eq:a}
\end{equation}
The inter-ring integrals are obtained via the a similar procedure.
In the following, we define $(\tau_{\chi}^{\parallel},\,\tau_{\bar{\chi}}^{\parallel})=(\tau^{\parallel},\,\bar{\tau}^{\parallel})$
and equally for $\tau_{\chi}^{\perp},v_{\chi},N(\epsilon_{k_{\chi}}),F_{\chi}$.
The full scattering operator is thus
\begin{align}
S^{\text{intra}}\left[f_{\mathbf{k}_{\chi}}\right]= & -2\pi n_{i}\,F_{\chi}\,N\,v\left[\cos\phi_{\mathbf{k}_{\chi}}\left(\tau^{\parallel}\int\frac{d\phi}{2\pi}\,\mathcal{T}_{\chi\chi}^{s}-\tau^{\parallel}\int\frac{d\phi}{2\pi}\cos\phi\,\mathcal{T}_{\chi\chi}^{s}-\tau^{\perp}\int\frac{d\phi}{2\pi}\,\sin\phi\,\mathcal{T}_{\chi\chi}^{a}\right)\right.\\
 & \left.+\sin\phi_{\mathbf{k}_{\chi}}\left(\tau^{\perp}\int\frac{d\phi}{2\pi}\mathcal{T}_{\chi\chi}^{s}-\tau^{\perp}\int\frac{d\phi}{2\pi}\cos\phi\mathcal{T}_{\chi\chi}^{s}+\tau^{\parallel}\int\frac{d\phi}{2\pi}\sin\phi\mathcal{T}_{\chi\chi}^{a}\right)\right]\\
S^{\text{inter}}\left[f_{\mathbf{k}_{\chi}}\right]= & -2\pi n_{i}F_{\chi}\,v\left[\cos\phi_{\boldsymbol{\mathbf{k}_{\chi}}}\left(\tau^{\parallel}\,N\,\int\frac{d\phi}{2\pi}\mathcal{T}_{\chi\bar{\chi}}^{s}-\bar{\tau}^{\parallel}\,\bar{N}\frac{\bar{v}}{v}\int\frac{d\phi}{2\pi}\cos\phi\mathcal{T}_{\bar{\chi}\chi}^{a}-\bar{\tau}^{\perp}\,\bar{N}\frac{\bar{v}}{v}\int\frac{d\phi}{2\pi}\sin\phi\,\mathcal{T}_{\bar{\chi}\chi}^{a}\right)\right.\\
 & \left.-\sin\phi_{\mathbf{k}_{\chi}}\left(\tau^{\perp}\,N\int\frac{d\phi}{2\pi}\mathcal{T}_{\chi\bar{\chi}}^{s}-\bar{\tau}^{\perp}\,\bar{N}\frac{\bar{v}}{v}\int\frac{d\phi}{2\pi}\cos\phi\mathcal{T}_{\bar{\chi}\chi}^{s}+\bar{\tau}^{\parallel}\bar{N}\frac{\bar{v}}{v}\int\frac{d\phi}{2\pi}\sin\phi\mathcal{T}_{\bar{\chi}\chi}^{a}\right)\right]\,.
\end{align}
Equating the coefficients of $\cos\phi_{\mathbf{k}},\,\sin\phi_{\mathbf{k}}$
on the LHS and RHS of the linearized BTEs, we obtain for the steady
state:
\begin{align}
-1 & =\tau^{\parallel}\left(\Gamma_{\chi\chi}^{0}-\Gamma_{\chi\chi}^{\text{cos}}+\Gamma_{\chi\bar{\chi}}^{0}\right)+\tau^{\perp}\,\Gamma_{\chi\chi}^{\text{sin}}-\bar{\tau}^{\parallel}\,\frac{\bar{v}}{v}\Gamma_{\bar{\chi}\chi}^{\text{cos}}+\bar{\tau}^{\perp}\,\frac{\bar{v}}{v}\Gamma_{\bar{\chi}\chi}^{\text{sin}}\,,\label{eq:eq1}\\
0 & =\tau^{\perp}\left(\Gamma_{\chi\chi}^{0}-\Gamma_{\chi\chi}^{\text{cos}}+\Gamma_{\chi\bar{\chi}}^{0}\right)-\tau^{\parallel}\,\Gamma_{\chi\chi}^{\text{sin}}-\bar{\tau}^{\perp}\,\frac{\bar{v}}{v}\Gamma_{\bar{\chi}\chi}^{\text{cos}}-\bar{\tau}^{\parallel}\,\frac{\bar{v}}{v}\Gamma_{\bar{\chi}\chi}^{\text{sin}}\,,\label{eq:eq2}
\end{align}
where, as defined already in the main text,
\begin{equation}
\Gamma_{\chi\bar{\chi}}^{(0,\cos,\sin)}=2\pi n_{i}N_{\chi}\int\frac{d\phi}{2\pi}\left\{ 1,\cos\phi,\sin\phi\right\} \left(\mathcal{T}_{\chi\bar{\chi}}^{s}(\phi)+\mathcal{T}_{\chi\bar{\chi}}^{a}(\phi)\right)\,.
\end{equation}
The system of equations can now be closed considering the respective
equations for the other channel $\bar{\chi}$, \emph{i.e.},
\begin{align}
-1 & =\bar{\tau}^{\parallel}\left(\Gamma_{\bar{\chi}\bar{\chi}}^{0}-\Gamma_{\bar{\chi}\bar{\chi}}^{\text{cos}}+\Gamma_{\bar{\chi}\chi}^{0}\right)+\bar{\tau}^{\perp}\,\Gamma_{\bar{\chi}\bar{\chi}}^{\text{sin}}-\tau^{\parallel}\,\frac{v}{\bar{v}}\Gamma_{\chi\bar{\chi}}^{\text{cos}}+\tau^{\perp}\,\frac{v}{\bar{v}}\Gamma_{\chi\bar{\chi}}^{\text{sin}}\,,\label{eq:eq3}\\
0 & =\bar{\tau}^{\perp}\left(\Gamma_{\bar{\chi}\bar{\chi}}^{0}-\Gamma_{\bar{\chi}\bar{\chi}}^{\text{cos}}+\Gamma_{\bar{\chi}\chi}^{0}\right)-\bar{\tau}^{\parallel}\,\Gamma_{\bar{\chi}\bar{\chi}}^{\text{sin}}-\tau^{\perp}\,\frac{v}{\bar{v}}\Gamma_{\chi\bar{\chi}}^{\text{cos}}-\tau^{\parallel}\,\frac{v}{\bar{v}}\Gamma_{\chi\bar{\chi}}^{\text{sin}}.\label{eq:eq4}
\end{align}
The four equations above can be manipulated by summing and subtracting
them to identify some common coefficient:
\begin{align}
-2 & =\tau^{\parallel}\Lambda_{-}+\bar{\tau}^{\parallel}\bar{\Lambda}_{-}+\tau^{\perp}\Upsilon_{+}+\bar{\tau}^{\perp}\bar{\Upsilon}_{+}\,,\\
0 & =\tau^{\parallel}\Lambda_{+}-\bar{\tau}^{\parallel}\bar{\Lambda}_{+}+\tau^{\perp}\Upsilon_{-}-\bar{\tau}^{\perp}\bar{\Upsilon}_{-}\,,\\
0 & =\tau^{\perp}\Lambda_{-}+\bar{\tau}^{\perp}\bar{\Lambda}_{-}-\tau^{\parallel}\Upsilon_{+}-\bar{\tau}^{\parallel}\bar{\Upsilon}_{+}\,,\\
0 & =\tau^{\perp}\Lambda_{+}-\bar{\tau}^{\perp}\bar{\Lambda}_{+}-\tau^{\parallel}\Upsilon_{-}+\bar{\tau}^{\parallel}\bar{\Upsilon}_{-}\,,
\end{align}
or in matrix form 
\begin{align}
\left(\begin{array}{cccc}
\Lambda_{-} & \bar{\Lambda}_{-} & \Upsilon_{+} & \bar{\Upsilon}_{+}\\
\Lambda_{+} & -\bar{\Lambda}_{+} & \Upsilon_{-} & -\bar{\Upsilon}_{-}\\
-\Upsilon_{+} & -\bar{\Upsilon}_{+} & \Lambda_{-} & \bar{\Lambda}_{-}\\
-\Upsilon_{-} & \bar{\Upsilon}_{-} & \Lambda_{+} & -\bar{\Lambda}_{+}
\end{array}\right)\left(\begin{array}{c}
\tau^{\parallel}\\
\bar{\tau}^{\parallel}\\
\tau^{\perp}\\
\bar{\tau}^{\perp}
\end{array}\right) & =\left(\begin{array}{c}
-2\\
0\\
0\\
0
\end{array}\right)\\
\Longleftrightarrow\left(\begin{array}{cc}
\hat{\Lambda} & \hat{\Upsilon}\\
-\hat{\Upsilon} & \hat{\Lambda}
\end{array}\right)\left(\begin{array}{c}
\mathbf{\boldsymbol{\tau}^{\parallel}}\\
\mathbf{\boldsymbol{\tau}^{\perp}}
\end{array}\right)= & -2\left(\begin{array}{c}
\mathbf{1}\\
\mathbf{0}
\end{array}\right)\,,\label{eq:CompactVersion}
\end{align}
where we have defined 
\begin{align}
\Lambda_{\pm} & =\Gamma_{\chi\chi}^{0}-\Gamma_{\chi\chi}^{\text{cos}}+\Gamma_{\chi\bar{\chi}}^{0}\pm\frac{v}{\bar{v}}\Gamma_{\chi\bar{\chi}}^{\text{\ensuremath{\cos}}}\,,\\
\Upsilon_{\pm} & =\Gamma_{\chi\chi}^{\text{sin}}\pm\frac{v}{\bar{v}}\Gamma_{\chi\bar{\chi}}^{\text{sin}}\,,
\end{align}
and analogously for their barred version, obtained from the last two
equations by replacing $\chi\to\bar{\chi}$. In our compact notation
we have $\left(\mathbf{\boldsymbol{\tau}^{\parallel}},\mathbf{\boldsymbol{\tau}^{\perp}}\right)^{\text{t}}=(\tau^{\parallel},\bar{\tau}^{\parallel},\tau^{\perp},\bar{\tau}^{\perp})^{\text{t}}\,,$
and $\left(\mathbf{1},\mathbf{0}\right)^{\text{t}}=(1,0,0,0)^{\text{t}}$.
Together with the corresponding system at \textbf{$K^{\prime}$ }valley
we thus identify 16 relaxation rates. In Fig.\,\ref{fig:Graphic-visualization-of}
we report a graphical visualization of the impurity scattering processes
and associated rates. Note that the number of relaxation rates doubles
when intervalley scattering processes are considered. 
The formal solution of the linear system Eq.\,\eqref{eq:CompactVersion}
gives
\begin{align}
\boldsymbol{\tau}^{\parallel} & =-2\left(\hat{\Lambda}+\hat{\Upsilon}\hat{\Lambda}^{-1}\hat{\Upsilon}\right)^{-1}\boldsymbol{1}\,,\label{eq:tauPar-1}\\
\boldsymbol{\tau}^{\perp} & =\hat{\Lambda}^{-1}\hat{\Upsilon}\boldsymbol{\tau}^{\parallel}\,,\label{eq:tauPer}
\end{align}
as reported in Eq.\,(6) of the main text. 

\subsubsection{Finite magnetic field\label{subsec:Finite-magnetic-field}}

In the presence of an external magnetic field, the LHS of the BTEs
reads as
\begin{equation}
\dot{\mathbf{k}}\cdot\nabla_{\mathbf{k}}f_{\mathbf{k}_{\chi}}=-\left(\boldsymbol{\mathcal{E}}+\text{\textbf{v}}_{\mathbf{k}}\times\boldsymbol{\mathcal{B}}\right)\cdot\nabla_{\mathbf{k}}\left(f_{\mathbf{k}_{\chi}}^{0}+\delta f_{\mathbf{k}_{\chi}}\right)\,.
\end{equation}
In the linear response regime, the contraction with the electric field
only selects the equilibrium part $f_{\mathbf{k}_{\chi}}^{0}$, as
seen above. On the other hand, the contraction with the magnetic field
selects the non-equilibrium part since
\begin{equation}
\left(\text{\textbf{v}}_{k}\times\boldsymbol{\mathcal{B}}\right)\cdot\text{\textbf{v}}_{k}=0\,.
\end{equation}
It is thus convenient to use the following generalized ansatz:
\begin{equation}
\delta f_{\mathbf{k}_{\chi}}=v_{\mathbf{k}_{\chi}}\left(\tau_{\chi}^{\parallel}\cos\phi_{\mathbf{k}}+\tau_{\chi}^{\perp}\sin\phi_{\mathbf{k}}\right)\,.\label{eq:NewAnsatz-1}
\end{equation}
In evaluating the term $\nabla_{\mathbf{k}}\delta f_{\mathbf{k}_{\chi}}$,
we use the relations between cartesian and polar derivates 
\begin{align}
 & \frac{\partial}{\partial k_{x}}=\cos\phi_{\mathbf{k}}\frac{\partial}{\partial k}-\frac{\sin\phi_{\mathbf{k}}}{k}\frac{\partial}{\partial\phi_{\mathbf{k}}}\,,\label{eq:PolarCartesian1}\\
 & \frac{\partial}{\partial k_{y}}=\sin\phi_{\mathbf{k}}\frac{\partial}{\partial k}+\frac{\cos\phi_{\mathbf{k}}}{k}\frac{\partial}{\partial\phi_{\mathbf{k}}}\,.\label{eq:PolarCartesian}
\end{align}
We thus have (omitting the index $\chi$ in the intermediate steps)
\begin{align}
\left(\text{\textbf{v}}_{\mathbf{k}}\times\boldsymbol{\mathcal{B}}\right)\cdot\nabla_{\mathbf{k}}\delta f_{\mathbf{k}} & =\epsilon_{abc}v_{\mathbf{k}}^{a}\mathcal{B}_{b}\,\partial_{c}\left[v_{\mathbf{k}}\left(\tau^{\parallel}\cos\phi_{\mathbf{k}}+\tau^{\perp}\sin\phi_{\mathbf{k}}\right)\right]\,,\label{eq:ba}
\end{align}
where $\epsilon_{abc}$ is the Levi-Civita symbol. Expanding the derivatives
(and using $\partial_{k_{i}}\to\partial_{i}$ for brevity), we obtain
\begin{align}
1.\quad\:\partial_{x}\left[v_{\mathbf{k}}\left(\tau^{\parallel}\cos\phi_{\mathbf{k}}+\tau_{\chi}^{\perp}\sin\phi_{\mathbf{k}}\right)\right] & =\cos^{2}\phi_{\mathbf{k}}\,\partial_{k}\left(v_{\mathbf{k}}\tau^{\parallel}\right)+\cos\phi_{\mathbf{k}}\sin\phi_{\mathbf{k}}\partial_{k}\left(v_{\mathbf{k}}\tau^{\perp}\right)+\nonumber \\
 & +\tau^{\parallel}\,v_{\mathbf{k}}\,\partial_{x}\cos\phi_{\mathbf{k}}+\tau^{\perp}v_{\mathbf{k}}\,\partial_{x}\sin\phi_{\mathbf{k}\,,}\\
2.\quad\:\partial_{y}\left[v_{\mathbf{k}}\left(\tau^{\parallel}\cos\phi_{\mathbf{k}}+\tau_{\chi}^{\perp}\sin\phi_{\mathbf{k}}\right)\right] & =\sin\phi_{\mathbf{k}}\cos\phi_{\mathbf{k}}\,\partial_{k}\left(v_{\mathbf{k}}\tau^{\parallel}\right)+\sin^{2}\phi_{\mathbf{k}}\,\partial_{y}\left(v_{\mathbf{k}}\tau^{\perp}\right)+\nonumber \\
 & +\tau^{\parallel}v_{\mathbf{k}}\,\partial_{y}\cos\phi_{\mathbf{k}}+\tau^{\perp}v_{\mathbf{k}}\,\partial_{y}\sin\phi_{\mathbf{k}\,,}
\end{align}
where we assumed an isotropic Fermi surface $\mathbf{k}\parallel\text{\textbf{v}}$.
These expressions, using Eqs.\,\eqref{eq:PolarCartesian1}-\eqref{eq:PolarCartesian},
can be rewritten as 
\begin{align}
1.\quad & \cos^{2}\phi_{\mathbf{k}}\,\partial_{k}\left(v_{\mathbf{k}}\tau^{\parallel}\right)+\cos\phi_{\mathbf{k}}\sin\phi_{\mathbf{k}}\,\partial_{k}\left(v_{\mathbf{k}}\tau^{\perp}\right)+v_{\mathbf{k}}\frac{\tau^{\parallel}}{k}\sin^{2}\phi_{k}-v_{\mathbf{k}}\frac{\tau^{\perp}}{k}\sin\phi_{\mathbf{k}}\cos\phi_{\mathbf{k}}\,,\label{eq:der1}\\
2.\quad & \sin\phi_{\mathbf{k}}\cos\phi_{\mathbf{k}}\,\partial_{k}\left(v_{\mathbf{k}}\tau^{\parallel}\right)+\sin^{2}\phi_{\mathbf{k}}\,\partial_{k}\left(v_{\mathbf{k}}\tau^{\perp}\right)-v_{\mathbf{k}}\frac{\tau^{\parallel}}{k}\sin\phi_{\mathbf{k}}\cos\phi_{\mathbf{k}}+v_{\mathbf{k}}\frac{\tau^{\perp}}{k}\cos^{2}\phi_{\mathbf{k}}\,.\label{eq:der2}
\end{align}
Taking a perpendicular magnetic field $\boldsymbol{\mathcal{B}=}\mathcal{B}\,\hat{z}$,
one obtains after standard algebraic manipulations
\begin{align}
\left(\text{\textbf{v}}_{\mathbf{k}_{\chi}}\times\boldsymbol{\mathcal{B}}\right)\cdot\nabla_{\mathbf{k}}\delta f_{\mathbf{k}_{\chi}} & =v_{\mathbf{k}_{\chi}}\omega_{\mathcal{B}}^{\chi}\left(\tau_{\chi}^{\parallel}\sin\phi_{\mathbf{k}_{\chi}-}\tau_{\chi}^{\perp}\cos\phi_{\mathbf{k}_{\chi}}\right)\,,
\end{align}
where we reinstated the index $\chi$ and defined the cyclotronic
frequency of the $\chi$-ring
\begin{equation}
\omega_{\mathcal{B}}^{\chi}=\frac{v_{\mathbf{k}_{\chi}}}{k_{\chi}}\mathcal{B}\,.\label{eq:CyclotronFreq}
\end{equation}
It is clear from Eq.\,\eqref{eq:eq1}-\eqref{eq:eq2}, that $\omega_{\mathcal{B}}^{\chi}$
can be reabsorbed in the definition of the skew cross sections:
\begin{equation}
\Gamma_{\chi\chi}^{\text{sin}}\to\Gamma_{\chi\chi}^{\text{sin}}+\omega_{B}^{\chi}\,,
\end{equation}
to which the trivial generalization of Eqs.\,\eqref{eq:CompactVersion}
follows. Note however, due to the slight different ansatz we have
used, the column of the know terms $-2\textrm{\ensuremath{\left(\boldsymbol{1},\boldsymbol{0}\right)}}^{\textrm{t}}$
has to be generalized to $-2F_{\chi}\left(\boldsymbol{1},\boldsymbol{0}\right)^{\textrm{t}}$.

\subsection{TRANSVERSE SCATTERING CROSS SECTION: ELASTIC AND SPIN-FLIP CHANNELS}

Below, we discuss the effective ``spin-Lorentz forces'' responsible
for the sign-change in the extrinsic transverse response and show
that the outer ring $\chi=1$ generally yields the dominant contribution
to the Hall conductivity in the weak scattering limit.

\subsubsection{Scattering in region I and II: intra-ring transition and main contribution
from $\chi=1$ ring}

\begin{figure}
\includegraphics[width=0.7\columnwidth]{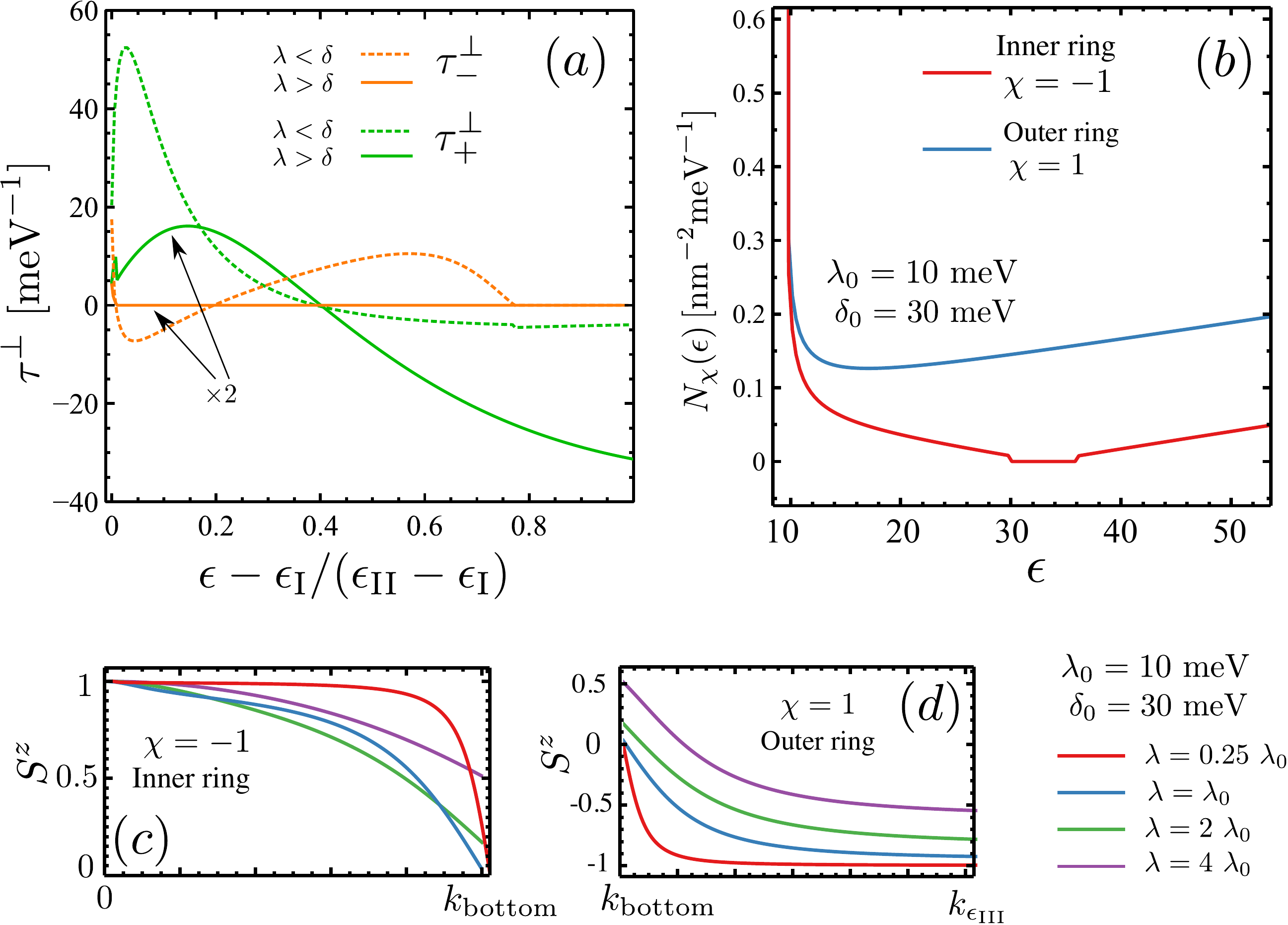}\caption{\label{fig:-in-two}(a) Transverse scattering time in two cases of
interest $\lambda\gtrless\delta$, where the larger and smaller energy
scales are, respectively, 10 and 30 meV. The change of sign is clearer
for $\tau_{+}^{\perp}$. (b) Also the density of states of the outer
ring is generally larger (here $\lambda>\delta$). (c) Finally it
is shown the sign change in the out-of-plane spin polarization happens
within the ring $\chi=1$. Parameters: $u_{0}=0.36$ eV$\cdot$nm$^{-2}$
and $n_{i}=10^{12}\,\text{cm}^{-2}$. }
\end{figure}

The starting point is the expression for the single-valley Hall conductivity
at $T=0$, 
\begin{equation}
\sigma_{\perp}=\frac{e^{2}}{h}\sum_{\chi=\pm1}k_{\chi}v_{\chi}\tau_{\chi}^{\perp}\,.
\end{equation}
Clearly, the peculiar sign-change must result from the energy dependence
of transverse scattering times $\tau_{\chi}^{\perp}$. In Fig.\,\eqref{fig:-in-two}
we show a comparison between the two $\tau_{\chi}^{\perp}$ where
inter-ring transition are neglected. Both $\tau_{\chi}^{\perp}$ change
sign, although $|\tau_{+}^{\perp}|>|\tau_{-}^{\perp}|$. The outer
ring is also associated with a larger density of states $N_{+}>N_{-}$
, where $N_{\chi}=k_{\chi}|2\pi v_{k_{\chi}}|^{-1}$, as displayed
in Fig.\,\ref{fig:-in-two}(b). The larger $\tau_{+}^{\perp}$ and
$N_{+}^{\perp}$ as compared to their $\chi=-1$ counterparts motivate
our discussion in the main text concerning the change of sign of $\sigma_{\perp}$
focused on $\text{\textbf{k}}_{+}\to\text{\textbf{k}}_{+}$ transitions.
In addition, the out-of-plane spin polarization $S^{z}$ also changes
sign within the $\chi=1$ ring, as displayed in Fig.\,\ref{fig:-in-two}(c)
and (d). 

\subsubsection{Spin-conserving and spin-flip Lorentz force from the collision integral}

\begin{figure}[t]
\includegraphics[width=0.6\columnwidth]{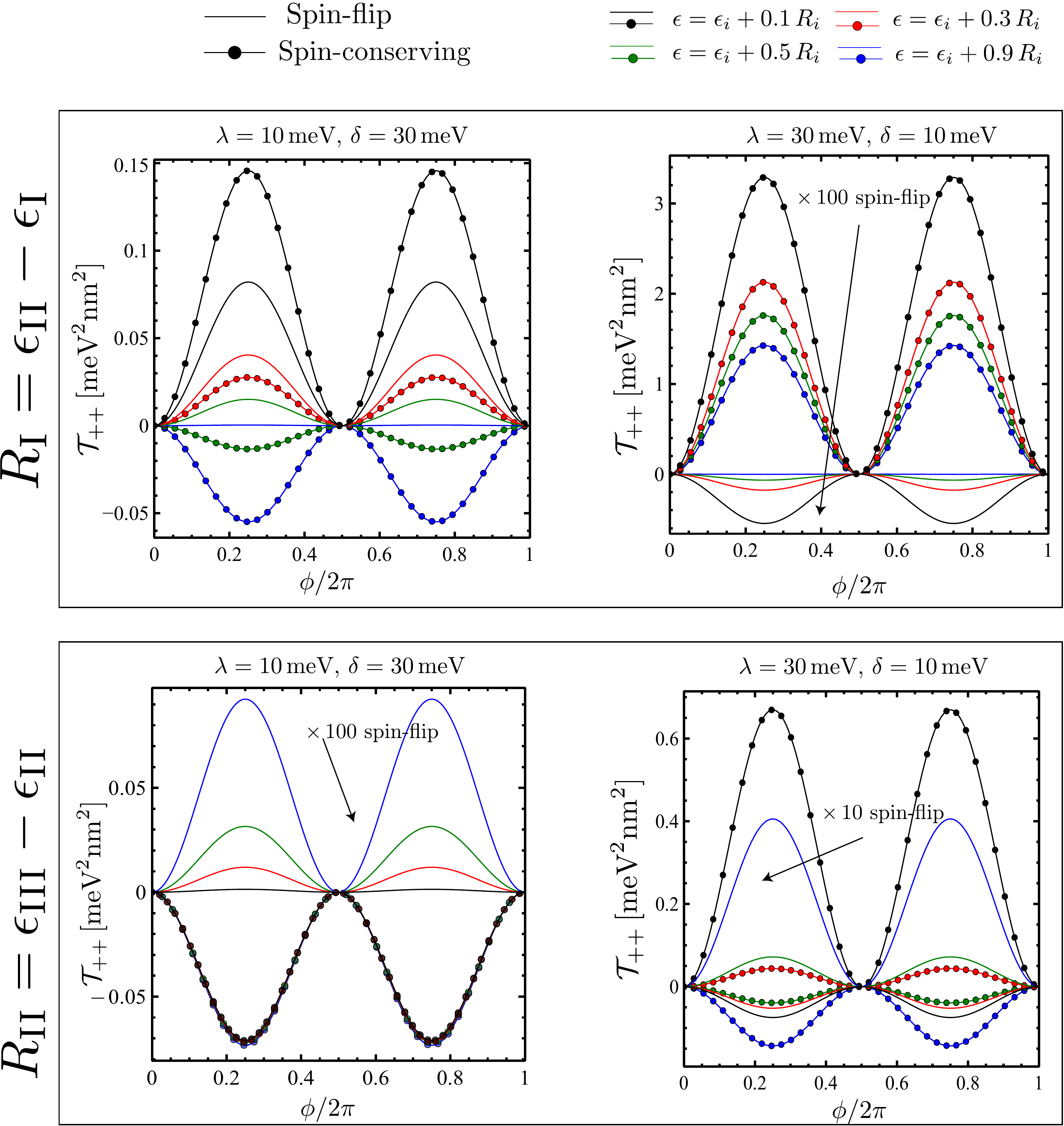}

\caption{\label{fig:Competition-between-the}Effective spin-conserving and
the spin-flip Lorentz forces. Parameters: $u_{0}=0.036$ eV$\cdot$nm$^{-2}$
and $n_{i}=10^{12}\,\text{cm}^{-2}$. }
\end{figure}

Having demonstrated the dominant role played by intra-ring $\chi$=1
processes, we now discuss the physical picture behind the sign-change
in $\sigma_{\perp}$ as the Fermi level approaches the spin majority
band edge. The transverse relaxation time, in the absence of a magnetic
field, is given by 
\begin{equation}
\tau_{+}^{\perp}=\frac{\Gamma_{++}^{\text{sin}}}{(\Gamma_{++}^{\text{0}}-\Gamma_{++}^{\cos})^{2}+(\Gamma_{++}^{\text{sin}})^{2}}\,,
\end{equation}
and hence the change of sign of $\tau_{+}^{\perp}$ is controlled
by the antisymmetric part of $\mathcal{T}_{++}(\phi)=|\langle\mathbf{k_{+}^{\prime}|}t|\mathbf{k}_{+}\rangle|^{2}.$
The single-impurity $T$ matrix can be decomposed according to the
following form 
\begin{equation}
t=t_{0}\gamma_{0}+t_{\text{KM}}\gamma_{\text{KM}}+t_{z}\,s_{z}+t_{m}\Sigma_{z}+t_{r}\gamma_{r}\,.\label{eq:Tmat_Structure}
\end{equation}
Importantly, all terms are associated with diagonal matrices in spin
space, except the ``Rashba term'' $T_{r}\gamma_{r}$. The latter
is indeed what connects orthogonal spin states. The resulting terms
in $\Gamma_{++}^{\sin}$ lead to the effective \emph{spin-conserving}
and \emph{spin-flip} Lorentz forces, as described in the main text.
In Fig.\,\ref{fig:Competition-between-the} we plot the modulus square
of $\mathcal{T}_{++}(\phi)$ as a function of the scattering angle
and for different values of the Fermi energy $\epsilon$ lying in
region I or II. 

\subsection{ADDITIONAL DISCUSSIONS }

\subsubsection{Thermal fluctuations}

The finite temperature transverse response is obtained from
\begin{equation}
\sigma_{\perp}^{T}(\mu)=-\int_{-\infty}^{+\infty}d\epsilon\,\left(\frac{\partial f^{0}(\epsilon,\mu,T)}{\partial\epsilon}\right)\sigma_{\perp}^{T=0}(\epsilon)\,,
\end{equation}
where $\sigma_{\perp}^{T=0}(\epsilon)$ is the zero-temperature response
and $f^{0}(\epsilon,\mu,T)=\left(1+\text{Exp}\left[\left(\epsilon-\mu\right)/k_{B}T\right]\right)^{-1}$.
Figure \,\eqref{fig:Effect-of-temperature} shows the temperature
dependence of the anomalous Hall conductivity for typical parameters.
The characteristic change of sign is visible up to temperatures $T\approx T_{\text{room}}/2$. 

\subsubsection{Intervalley scattering}

\begin{figure}
\centering{}\includegraphics[width=0.6\columnwidth]{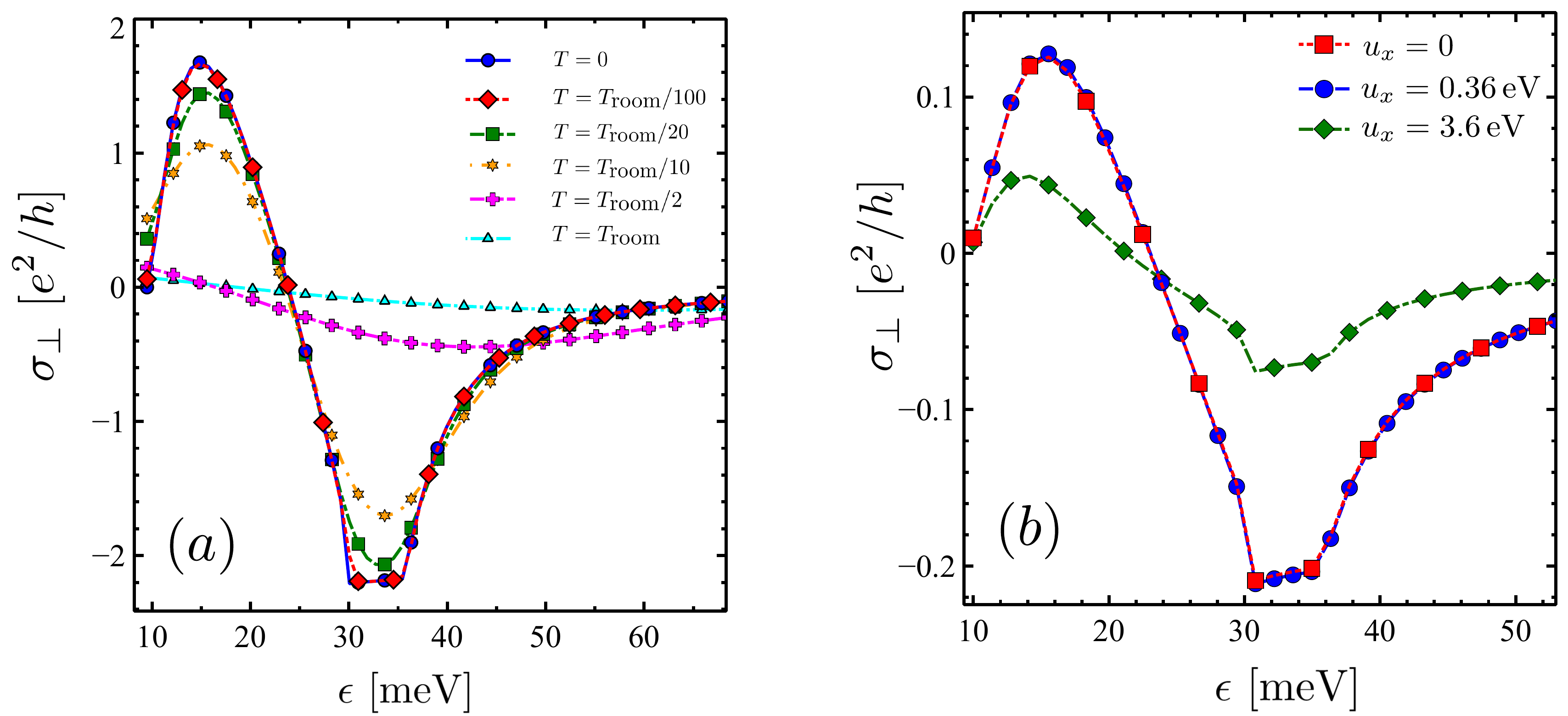}\caption{Effect of thermal fluctuations and intervalley scattering. (a) Change
of sign in $\sigma_{\perp}$ is distinguishable for $T\lesssim T_{\text{room}}/2$,
where $T_{\text{room}}=300\,\text{K}$. \label{fig:Effect-of-temperature}
(b) In the less favorable scenario ($u_{x}=u_{0}$), the transport
times $\tau^{\parallel,\perp}(\epsilon)$ are reduced by a factor
$\approx2$. Parameters: $u_{0}=3.6$ eV$\cdot$nm$^{-2}$ and $n=10^{12}\,\text{cm}^{-2}$.}
\end{figure}

To account for intervalley scattering, we use the following simplified
model for the disorder potential with site $C_{6v}$ symmetry: 
\begin{equation}
V\left(\text{\textbf{x}}\right)=\sum_{i=1}^{N_{i}}\,R^{2}(u_{0}\gamma_{0}+u_{x}\tau_{x})\delta(\text{\textbf{x}}-\boldsymbol{\text{\textbf{x}}}_{i})\,,
\end{equation}
which interpolates between a pure intravalley potential $(u_{x}=0)$
and a atomically-sharp potential with $u_{x}=u_{0}$ (e.g., a resonant
adatom) leading to strong intervalley scattering \cite{Impurities_G_Basko,supp2_Pachoud}.
The intervalley term ($u_{x}$) introduces new rates:
\begin{equation}
\boldsymbol{\Gamma}_{\chi\zeta}=\left\{ \Gamma_{\chi\zeta}^{0},\Gamma_{\chi\zeta}^{\cos},\Gamma_{\chi\zeta}^{\text{sin}}\right\} =2\pi n_{i}\int(d^{2}\text{\textbf{k}}^{\prime})\,\left\{ 1,\cos\phi,\sin\phi\right\} \mathcal{T}_{\chi\zeta}(\phi)\,,
\end{equation}
where $\zeta=\pm1$ is an index associated with outer and inner ring
respectively for states at $K^{\prime}$ valley, obtainable from Eq.\,(3)
of the main text by performing the substitution $\lambda,v\to-\lambda,-v$.
We plot the result in Fig.\,\ref{fig:Effect-of-temperature}. The
intervalley scattering results in a reduction of the total transverse
relaxation time. 

\subsubsection{Magnetic field}

\begin{figure}
\centering{}\includegraphics[width=0.4\columnwidth]{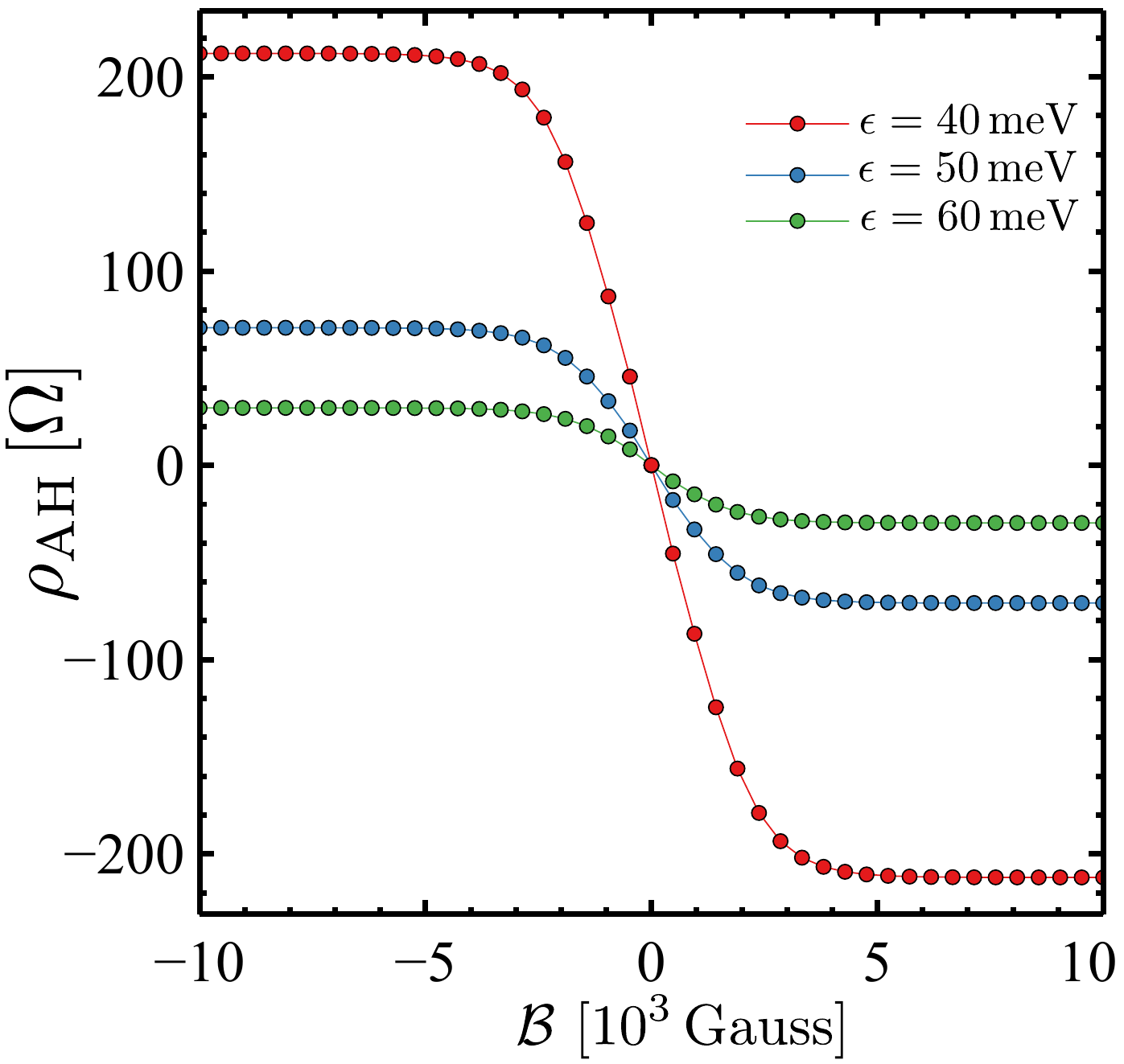}\caption{$\rho_{\text{AH}}$ as a function of the magnetic field for different
Fermi energies $\epsilon$. Parameters: $\lambda=10\,\text{meV},$
$u_{0}=1$ eV$\cdot$nm$^{-2}$ and $n_{i}=10^{12}$. Also $\delta_{\text{eff}}=\delta\,\text{tanh}\left(\mathcal{B}/\mathcal{B}_{\text{sat}}\right)$,
$\delta=30$ meV, and $\mathcal{B}_{\text{sat}}=1.5\times10{}^{3}\text{Gauss}$
\cite{AHE_G_Tang_18}.\label{fig:-as-a-1}}
\end{figure}

The formalism at finite magnetic field {[}Sec.~\eqref{subsec:Finite-magnetic-field}{]}
allows us to model various quantities of interest in graphene/thin
film heterostructures. Figure\,\ref{fig:-as-a-1} shows the anomalous
Hall resistivity $\rho_{\text{AH}}$ as function of the applied magnetic
field. To reproduce a typical hysteresis loop for magnetized graphene,
the effective exchange coupling is chosen as $\delta_{\text{eff}}=\delta\,\text{tanh}\left(\mathcal{B}/\mathcal{B}_{\text{sat}}\right)$,
where $\delta=30$ meV, and with the saturation field $\mathcal{B}_{\text{sat}}=1.5\times10{}^{3}\text{Gauss}$
\cite{Ferr_G_YIG_Wang15,AHE_G_Tang_18}. 

\section{EQUIVALENCE BETWEEN SEMICLASSICAL AND KUBO-STREDA FORMALISM\label{sec:SKEW-SCATTERING-CONTRIBUTION:}}

We evaluate the disorder correction to the anomalous Hall conductivity
by means of linear response theory \cite{supp_2bStreda,supp3_Crepieux,supp4_Sinitsyn,Ado2015,milletari16}.
The results are shown to agree with the BTEs to numerical accuracy.

\subsection{Linear response formalism}

The longitudinal (Drude) and transverse (Hall) responses are denotes
as $\sigma_{xx}$ and $\sigma_{yx}$, respectively. The transverse
response is calculated using the Kubo-Streda formalism \cite{supp_2bStreda},
$\sigma_{yx}(\epsilon)=\sigma_{yx}^{I}(\epsilon)+\sigma_{yx}^{II}(\epsilon)\,,$
where \cite{Comment2}
\begin{align}
\sigma_{yx}^{I}(\epsilon) & =\int(d\mathbf{k})\,\text{tr}\left(v_{y}\mathcal{G}_{\mathbf{k}}^{R}(\epsilon)\tilde{v}_{x}(\epsilon)\mathcal{G}_{\mathbf{k}}^{A}(\epsilon)\right)\,,\label{eq:sigma1}\\
\sigma_{yx}^{II}(\epsilon) & =-\int_{-\infty}^{\epsilon}d\omega\int(d\mathbf{k})\,\text{Re}\left(\text{tr}\left(v_{x}\frac{d\mathcal{G}_{\mathbf{k}}^{R}(\epsilon)}{d\epsilon}v_{y}\mathcal{G}_{\mathbf{k}}^{R}(\epsilon)-v_{x}\mathcal{G}_{\mathbf{k}}^{R}(\epsilon)v_{y}\frac{d\mathcal{G}_{\mathbf{k}}^{R}(\epsilon)}{d\epsilon}\right)\right)\,,\label{eq:sigma2}
\end{align}
and
\begin{equation}
\mathcal{G}_{\mathbf{k}}^{R/A}(\epsilon)=\left(\epsilon-H_{0\mathbf{k}}-\Sigma^{R/A}\right)^{-1}\,,\label{eq:avGF}
\end{equation}
are disorder averaged Green's functions associated with the Hamiltonian
Eq.\,(1) of the main text and $\Sigma^{R/A}=n_{i}t^{R/A}$ is the
self energy. Here, $\text{tr}$ is the trace over internal degrees
of freedom $\tau,\Sigma,s$ and $v_{x,y}=\partial_{k_{x,y}}H_{0\boldsymbol{k}}=\tau_{z}\sigma_{x,y}$
are the bare current operators. $\tilde{v}_{x}$ is the renormalized
current vertex, obtained from the Bethe-Salpeter (BS) equation
\begin{equation}
\tilde{v}_{x}(\epsilon)=v_{x}+n_{i}\,\text{tr}\,\sum_{\mathbf{k}}t^{R}(\epsilon)\,\mathcal{G}_{\mathbf{k}}^{R}(\epsilon)\,\tilde{v}_{x}(\epsilon)\,t^{A}(\epsilon)\,\mathcal{G}_{\mathbf{k}}^{A}(\epsilon)\,.\label{eq:BSvertex}
\end{equation}
$\sigma_{yx}^{I},\sigma_{yx}^{II}$ are the so-called ``Fermi surface''
and ``Fermi sea'' contribution to the Hall response . The semiclassical
(skew scattering) contribution originates from the leading disorder
correction, $\sigma_{\perp}\propto\tau^{\perp}\propto n_{i}^{-1}$
\cite{milletari16}. 

\begin{figure}
\centering{}\includegraphics[width=0.5\columnwidth]{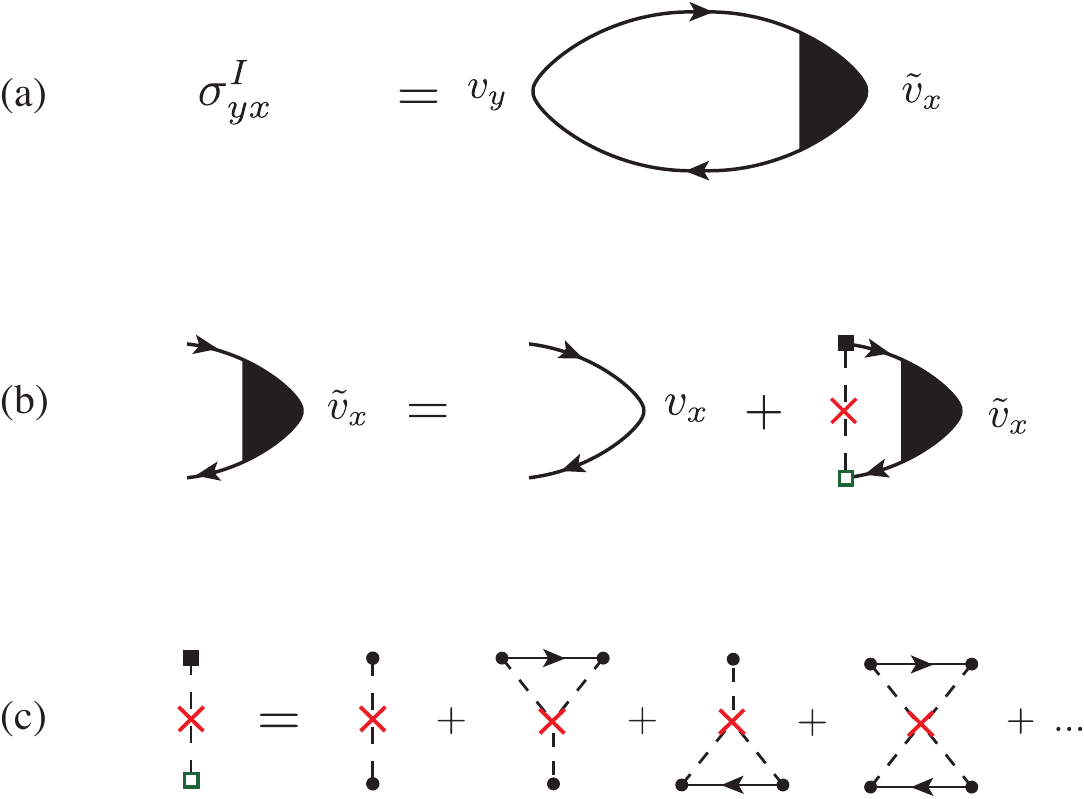}\caption{$T$-matrix ladder resummation for the renormalized charge vertex
entering the transverse response \cite{milletari16}. (a) the solid
line with arrow towards right (left) represents a renormalized propagator
projected on the retarded (advanced) sector. (b) diagrammatic representation
of the Bethe-Salpeter $T$-matrix ladder, as shown in (c). Black (white)
boxes correspond to $T^{R}$($T^{A}$), red crosses to impurity density
insertions and black dot to single impurity scattering potential $u_{0}$.
\label{fig:-matrix-ladder-resummation}}
\end{figure}

\subsubsection{Disorder averaged Green's function}

We provide the explicit form of the disorder averaged Green's function
for intravalley scattering potentials ($u_{x}=0$). To obtain those,
one first needs to calculate the disorder averaged $T$ matrix 
\begin{equation}
t^{R/A}=\langle V(1-g_{0}^{R/A}V)^{-1}\rangle_{\text{dis}}\,,\label{eq:tm}
\end{equation}
where the momentum-integrated bare Green's function reads as
\begin{align}
g_{0}^{R,A} & =-\frac{1}{2}\left\{ \left(\epsilon\,\gamma_{0}-\delta\,s_{z}\right)L_{+}^{R,A}+\frac{1}{D}\left[\epsilon\left(\lambda^{2}+\delta^{2}\right)\gamma_{0}-\delta\left(\epsilon^{2}-\lambda^{2}\right)\,s_{z}+\delta\,\lambda^{2}\,\gamma_{m}+\epsilon\,\lambda^{2}\gamma_{\text{KM}}-\lambda\left(\epsilon^{2}-\delta^{2}\right)\gamma_{r}\right]L_{-}^{R,A}\right\} \,,\label{eq:Gbare}
\end{align}
with $D=\sqrt{\epsilon^{2}\delta^{2}+\lambda^{2}\left(\epsilon^{2}-\delta^{2}\right)}$
. Also we have defined $L_{\pm}^{p}=L_{1}^{p}\pm L_{2}^{p}$ ($p=R/A=\pm$)
with
\begin{align}
\\
L_{1,2}^{p} & =\frac{1}{v^{2}k^{2}-z_{1,2}^{p}}\,,\;\;\;z_{1,2}^{p}=\epsilon^{2}+\delta^{2}\pm2\,D+i\,p\,0^{+}\,.
\end{align}
Beyond the identity and the matrix structures $\gamma_{r},s_{z}$,
already present in the bare Hamiltonian, we obtain two additional
terms $\gamma_{m}=\Sigma_{z},\,\gamma_{\text{KM}}=\Sigma_{z}s_{z}$,
which will appear in the self-energy and thus in the disorder averaged
Green's functions. 

To obtain an analytical expression for $\mathcal{G}_{\mathbf{k}}^{R/A}(\epsilon)$,
we consider an effective model containing \emph{all} the matrix structures
appearing in $g_{0}^{R/A}$, namely
\begin{equation}
H_{\text{eff}}=H_{0}+m\,\gamma_{\text{KM}}+\lambda_{m}\gamma_{m}\,.\label{eq:Heff}
\end{equation}
The disorder averaged Green's function  of the original Hamiltonian
Eq.\,(1) in the main text for $u_{x}=0$ is obtained by taking the
\emph{bare }Green's function of Eq.\,\eqref{eq:Heff} and\emph{ }performing
the following analytical continuations
\begin{eqnarray}
\mathcal{A}^{R/A}:\:\epsilon & \to & \epsilon-n_{i}(\epsilon\pm i\eta_{0})\,,\label{eq:substitution}\\
\lambda & \to & \lambda+n_{i}\left(\varDelta\lambda\pm i\eta_{r}\right)\,,\nonumber \\
\delta & \to & \delta+n_{i}\left(\varDelta\delta\pm i\eta_{z}\right)\,,\\
m_{0} & \to & n_{i}(\varDelta m\pm i\eta_{\text{KM}})\,,\nonumber \\
\lambda_{m} & \to & n_{i}\left(\varDelta\lambda_{m}\pm i\eta_{m}\right),
\end{eqnarray}
where the parameters defined by 
\begin{equation}
\varDelta\lambda_{i}+i\eta_{i}=\frac{1}{8}\text{tr}[t^{R}\gamma_{i}]\,,\label{eq:parameters}
\end{equation}
and $\lambda_{i},\eta_{i}$ denote, respectively, the real and imaginary
parts of one of the parts of the $T$ matrix associated with the matrix
$\gamma_{i}$. We then find
\begin{align}
\mathcal{G}^{p} & =\left.\sum_{\xi=\pm}\left(\mathcal{M}_{0,\xi}+\mathcal{M}_{\phi_{\text{\textbf{k}}},\xi}\right)\mathcal{L}_{\xi}\right|_{\mathcal{A}^{p}}\,,
\end{align}
meaning that the analytical continuation $\mathcal{A}^{p}$ has to
be performed to find the $p=R/A$ sectors. Above, expressing the matrices
in the notation $\mathcal{M}=\tau_{i}\otimes\mathcal{\tilde{M}}\,,\mathcal{\tilde{M}}_{jk}=1/4\,\text{tr}\left[\mathcal{\tilde{\mathcal{M}}}\sigma_{j}s_{k}\right]$,
we have explicitly 

\begin{align}
\mathcal{M}_{0,+}= & -\frac{\tau_{0}}{2}\left(\begin{array}{cccc}
\epsilon & 0 & 0 & -\delta\\
0 & 0 & 0 & 0\\
0 & 0 & 0 & 0\\
\lambda_{m} & 0 & 0 & m
\end{array}\right)\,,\\
\nonumber \\
\mathcal{M}_{0,-}= & -\frac{\tau_{0}}{2\tilde{\Gamma}}\left(\begin{array}{cccc}
\left[\epsilon\left(\lambda^{2}+\delta^{2}\right)+m\left(\lambda_{m}\,\delta-\lambda^{2}\right)\right] & 0 & 0 & \left[\delta^{2}\left(\lambda_{m}+\delta\right)-\epsilon\left(m\,\lambda_{m}+\epsilon\,\delta\right)\right]\\
0 & 0 & 0 & 0\\
0 & 0 & 0 & 0\\
\left[\lambda^{2}\left(\lambda_{m}+\delta\right)-m\left(\epsilon\,\delta+m\,\lambda_{m}\right)\right] & 0 & 0 & \left[\epsilon\left(\lambda^{2}-\delta\,\lambda_{m}\right)-m\left(\lambda^{2}+\lambda_{m}^{2}\right)\right]
\end{array}\right)\nonumber \\
 & -\frac{\delta}{2\tilde{\Gamma}}\left[\left(\lambda_{m}+\delta\right)^{2}-\left(\epsilon-m\right)^{2}\right]\gamma_{r}\,,
\end{align}
and
\begin{align}
\mathcal{M}_{\phi_{\text{\textbf{k}}},+} & =-\frac{\tau_{z}}{2}\left(\begin{array}{cccc}
0 & 0 & 0 & 0\\
vk\,\cos\phi_{\text{\textbf{k}}} & \lambda\sin2\phi_{\text{\textbf{k}}} & -\lambda\cos2\phi_{\text{\textbf{k}}} & 0\\
vk\,\sin\phi_{\text{\textbf{k}}} & -\lambda\cos2\phi_{\text{\textbf{k}}} & -\lambda\sin2\phi_{\text{\textbf{k}}} & 0\\
0 & 0 & 0 & 0
\end{array}\right)\\
\mathcal{M}_{\phi_{\text{\textbf{k}}},-} & =-\frac{\tau_{0}}{2\tilde{\Gamma}}\left(\begin{array}{cccc}
0 & vk\left(\epsilon-m\right)\lambda\,\sin\phi_{\text{\textbf{k}}} & -vk\left(\epsilon-m\right)\lambda\,\cos\phi_{\text{\textbf{k}}} & 0\\
0 & 0 & 0 & 0\\
0 & 0 & 0 & 0\\
0 & vk\,\lambda\left(\lambda_{m}+\delta\right)\sin\phi_{\text{\textbf{k}}} & -vk\,\lambda\left(\lambda_{m}+\delta\right)\cos\phi_{\text{\textbf{k}}} & 0
\end{array}\right)\nonumber \\
 & -\frac{\tau_{z}}{4\tilde{\Gamma}}\left(\begin{array}{cccc}
0 & 0 & 0 & 0\\
0 & \lambda\left(\epsilon^{2}+\delta^{2}-m^{2}-\lambda_{m}\right)\,\sin2\phi_{\text{\textbf{k}}} & -\lambda\left(\epsilon^{2}+\delta^{2}-m^{2}-\lambda_{m}\right)\,\cos2\phi_{\text{\textbf{k}}} & -vk\,(m\lambda_{m}+\epsilon\delta)\cos\phi_{\text{\textbf{k}}}\\
0 & -\lambda\left(\epsilon^{2}+\delta^{2}-m^{2}-\lambda_{m}\right)\,\cos2\phi_{\text{\textbf{k}}} & -\lambda\left(\epsilon^{2}+\delta^{2}-m^{2}-\lambda_{m}\right)\,\sin2\phi_{\text{\textbf{k}}} & -vk\,(m\lambda_{m}+\epsilon\delta)\sin\phi_{\text{\textbf{k}}}\\
0 & 0 & 0 & 0
\end{array}\right)
\end{align}
with 
\begin{align}
\tilde{\Gamma} & =\sqrt{\delta^{2}\left[\left(m-\epsilon\right)^{2}-\left(\lambda_{m}+\delta\right)^{2}\right]+\left(m\,\lambda_{m}+\epsilon\,\delta\right)^{2}}\\
\mathcal{L}_{\pm} & =\left[v^{2}k^{2}-\left(\epsilon^{2}+\delta^{2}-m^{2}-\lambda_{m}^{2}\pm2\tilde{\Gamma}\right)\right]^{-1}\,.
\end{align}

\subsubsection{Renormalized charge current vertex: solution of the BS equation and
symmetry arguments}

\begin{figure}
\includegraphics[width=0.7\columnwidth]{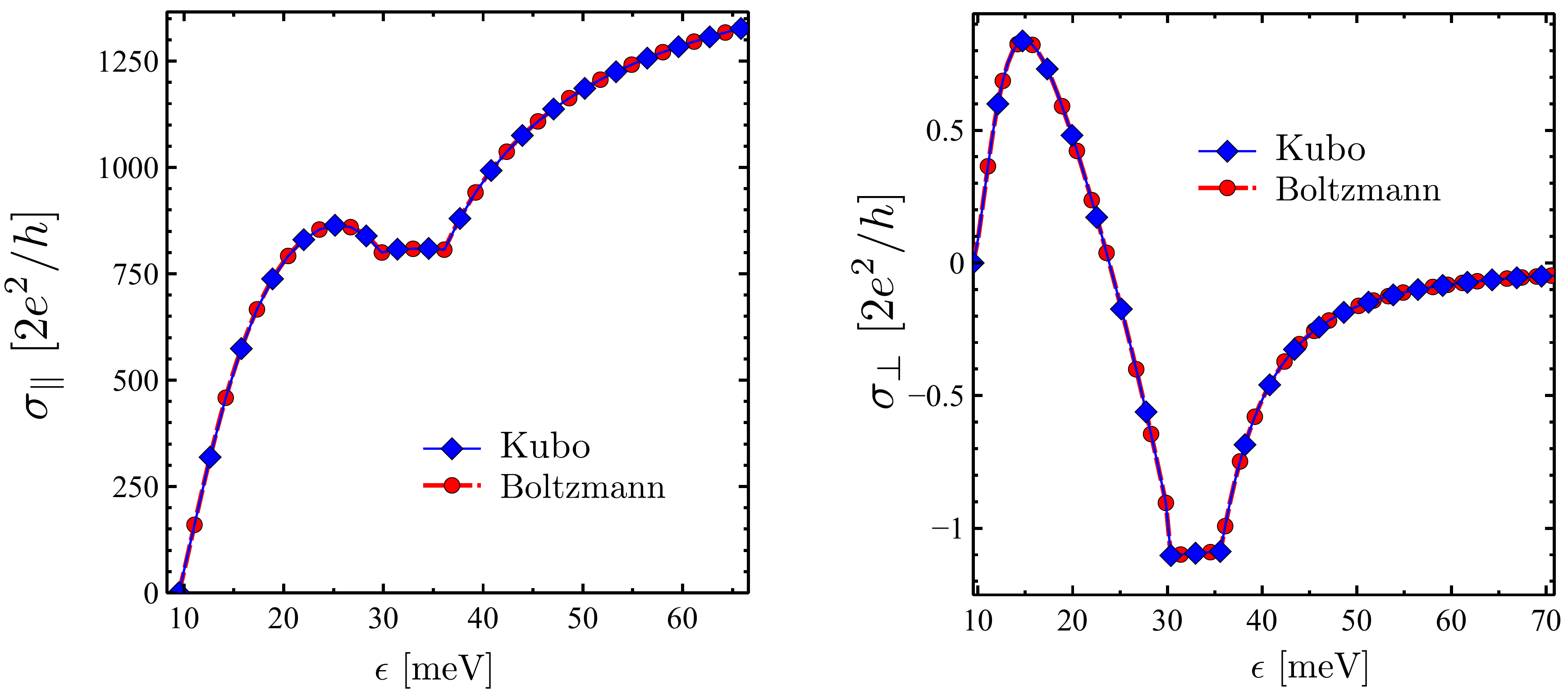}\caption{Comparison between Kubo\textendash Streda (diagrammatic) and Boltzmann
calculation for both the longitudinal and transverse response in a
representative case: $u_{0}=0.36$ eV$\cdot$nm$^{-2}$ a d $n_{i}=10^{12}\text{cm}^{-2}$.
The agreement between the two formalisms is excellent.\label{fig:Comparison-between-Kubo}}
\end{figure}
The recursive BS equation Eq.\,\eqref{eq:BSvertex} for the charge
vertex produces new matrix structures, which are associated to observables
that can be ``excited'' upon the application of external fields.
 We show now how these observables can be predicted by simple symmetry
arguments. The continuum model Eq.\,(1) of the main text is invariant
under the following symmetry operations
\begin{eqnarray}
C_{2} & : & \tau_{x}s_{z}e^{i\pi l_{z}}\label{eq:symmetries}\\
\Lambda_{z} & : & \tau_{z}\\
\Lambda_{x} & : & \tau_{x}\Sigma_{z}\\
\Lambda_{y} & : & \tau_{y}\Sigma_{z\,.}\label{eq:symmetries_2}
\end{eqnarray}
$C_{2}$ is the rotation of $\pi$ around the $\hat{z}$-axis, and
$l_{z}=-i\partial_{\phi}$ the generator of the rotation in coordinate
space. Reflection around $\hat{x}$: $R_{x}:\,\Sigma_{x}s_{y}r_{x}$,
with $r_{x}:\;\;(\mathbf{x},\mathbf{y})\to(\mathbf{x},-\mathbf{y})\,,$
and time reversal operation $\ensuremath{\Theta}$, present in the
continuum limit of bare graphene, are broken symmetries in this model;
we can refer to the transformations above as \emph{pseudosymmetries
\begin{align}
\text{\ensuremath{\Theta}}^{-1}\,H\left(\delta\right)\,\Theta & =H\left(-\delta\right)\\
R_{x}^{-1}\,H\left(\delta\right)\,R_{x} & =H\left(-\delta\right)\,.
\end{align}
}To constraint the number of observables with nonzero expectation
value, we examine the generic response function 
\begin{equation}
\mathcal{R}_{\alpha}=\text{Tr\ensuremath{\left[\gamma_{\alpha}\,G_{0}^{R}v_{x}\,G_{0}^{A}\right]}\,,}\label{eq:bubble}
\end{equation}
for an observable $\gamma_{\alpha}$. In the above, Tr includes the
trace in momentum space. Exploring any of the symmetries $\mathcal{S}$
listed in Eqs.\,\eqref{eq:symmetries}-\eqref{eq:symmetries_2}
\begin{equation}
\mathcal{S}:\,\,G_{0}^{R,A}\to G_{0}^{R,A}\,.
\end{equation}
Eq.\,\eqref{eq:bubble} can be manipulated as follows
\begin{align}
\mathcal{R}_{\alpha} & =\text{Tr}\ensuremath{\left[\left(\mathcal{S}\gamma_{\alpha}\mathcal{S}^{\dagger}\right)\left(\mathcal{S}\,G_{0}^{R}\mathcal{S}^{\dagger}\right)\left(\mathcal{S}v_{x}\mathcal{S}^{\dagger}\right)\left(\mathcal{S}\,G_{0}^{A}\mathcal{S}^{\dagger}\right)\right]}\\
 & =\epsilon_{x}\epsilon_{\alpha}\text{Tr}\ensuremath{\left[\gamma_{\alpha}\,G^{R}v_{x}\,G^{A}\right]}\,,
\end{align}
where $\epsilon=\pm1$ is the parity of some operator under $\mathcal{S}$.
From the last equation, we see that a non-zero response requires the
operator $\gamma_{\alpha}$ to have the same parity of the current
vertex $v_{x}$ under the action of $\mathcal{S}$. We have
\begin{eqnarray}
C_{2} & : & v_{x}\to-v_{x}\,,\\
\Lambda_{x,y,z} & : & v_{x}\to+v_{x}\,.
\end{eqnarray}
By spanning the 64-dimensional algebra $\tau_{i}\sigma_{j}s_{k}$,
we conclude that only eight responses are allowed 
\begin{eqnarray}
\tau_{0}\Sigma_{0}s_{x,y} & : & \text{spin-x,y (spin-Galvanic) magnetisation}\\
\tau_{0}\Sigma_{z}s_{x,y} & : & \text{staggered spin-x,y polarization}\\
\tau_{z}\Sigma_{x,}s_{0,z} & : & \text{x-charge current (Drude) and spin-z x-current}\\
\tau_{z}\Sigma_{y}s_{0,z} & : & \text{charge y-current (Hall) and z-spin y-current (spin Hall)}\,.
\end{eqnarray}
We verified this is confirmed by explicit diagrammatic calculation.
Solution of Eq.\,\eqref{eq:BSvertex} then requires the inversion
of a $8\times8$ matrix. The solution can be written in the form 
\begin{equation}
\tilde{v}_{x}=\sum_{i}v_{x}^{i}\gamma_{i}\,,\label{eq:solVer}
\end{equation}
where $\gamma_{i}$ as given above. By inserting the renormalized
vertex in Eq.\,\eqref{eq:solVer} into Eq.\,\eqref{eq:sigma1},
one can finally find the skew-scattering contribution to the Hall
response. One can also calculate the Drude response (xx) by replacing
$v_{y}\to v_{x}$ in Eq.\,\eqref{eq:sigma1}. Figure\,\ref{fig:Comparison-between-Kubo}
benchmarks the BTEs against the diagrammatic formalism. 

\section{SIDE-JUMP CONTRIBUTION}

\begin{figure}
\includegraphics[width=0.4\columnwidth]{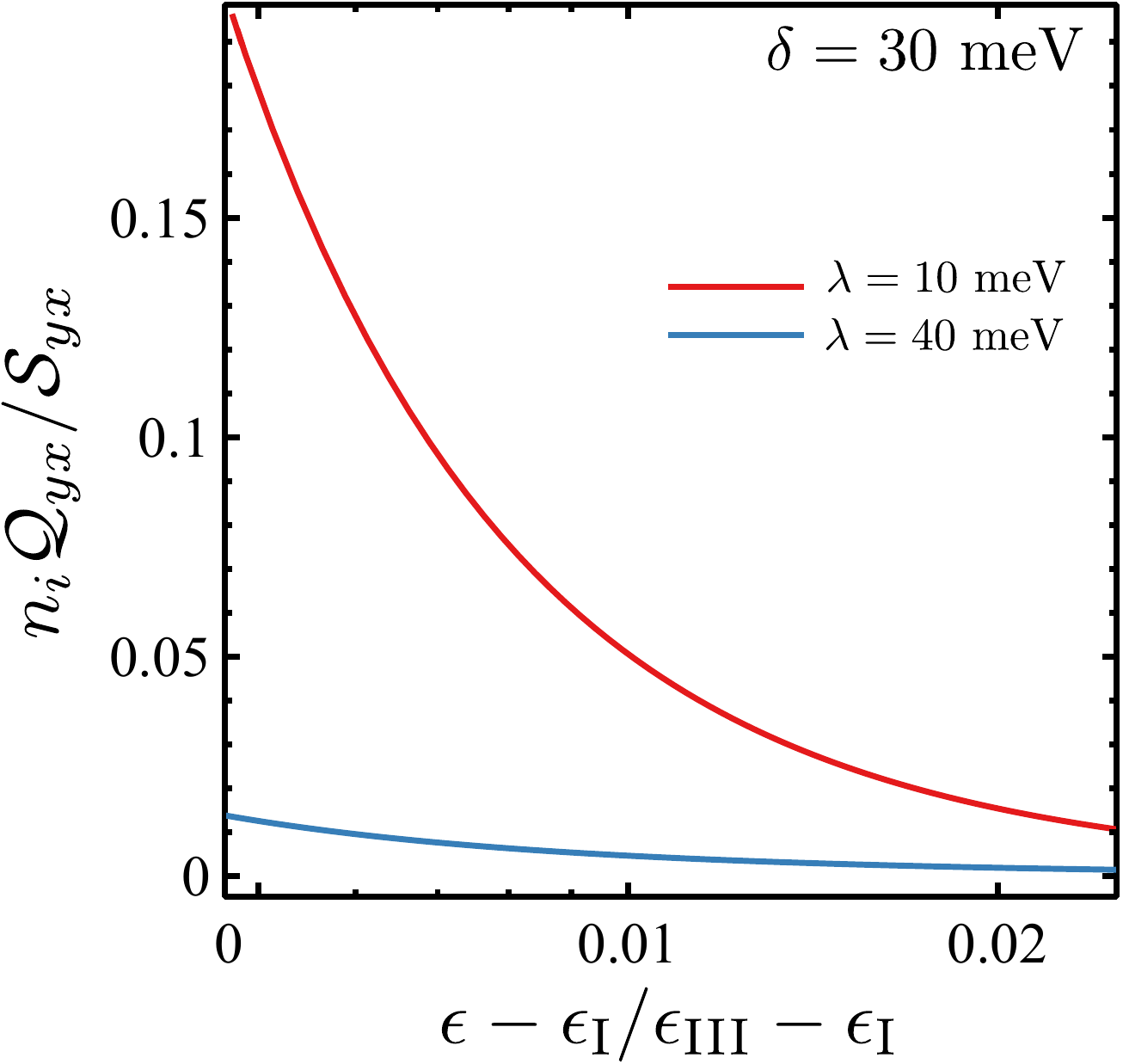}\caption{Ratio of the side-jump and skew scattering contribution to $\sigma_{yx}^{I}$
in high mobility samples. The side-jump contribution is negligible
away from the Dirac point $k_{F}l\gg1$. Parameters: $\lambda=$10
meV (red line) and $\lambda=40$ meV (blue line), $\lambda_{z}=30$
meV, $u_{0}=1$ eV$\cdot$nm$^{-2}$ and $n_{i}=10^{10}$ cm$^{-2}$.
\label{fig:Ratio-of-the}}
\end{figure}

We now evaluate the quantum side-jump (anomalous) contribution to
the transverse transport coefficients and show it provides a small
correction to the transverse response in clean samples with $k_{F}l\gg1$.
For brevity, we focus here on the AH response. The side-jump contribution
$\mathcal{Q}_{yx}$ is obtained by isolating the impurity concentration-independent
term 
\begin{equation}
\sigma_{yx}^{I}=\frac{\mathcal{S}_{yx}}{n_{i}}+\mathcal{Q}_{yx}+O(n_{i})\,.
\end{equation}
Within the rigorous diagrammatic formalism, this requires the calculation
of the ladder series for the renormalized vertex (see Fig.\,\ref{fig:-matrix-ladder-resummation}).
In Fig.\,\,\ref{fig:Ratio-of-the} we plot the ratio $\mathcal{S}_{yx}/n_{i}\mathcal{Q}_{yx}$
at selected SOC values in a system with high mobility. The results
show that the side-jump contribution is only sizable in a very narrow
energy window, i.e. \textasciitilde 1-10\% of the distance between
the Rashba edge $\epsilon_{\text{III}}$ and the bottom of the skyrmionic
band $\epsilon_{\text{I}}$. The side-jump part rapidly decays to
zero, in the diffusive regime with $k_{F}l\gg1$, thus justifying
our approximation in the main text $\sigma_{yx}^{I}\simeq\mathcal{S}_{yx}/n_{i}$.
We note that the anomalous term $\mathcal{Q}_{yx}$ also receives
contribution from the so-called $\Psi$ and $X$ diagrams encoding
quantum coherent skew scattering not evaluated here (see \cite{milletari16}
for more details). 

\section{INTRINSIC CONTRIBUTION TO THE AHE}

In this section we give details about the calculation of the intrinsic
contribution to the AHE. This can be done in two equivalent ways:
via a direct Berry-curvature calculation or using the clean limit
of the Kubo-Streda formula. We show they provide the same result. 

\subsubsection{Kubo-Streda approach}

We start here by computing the intrinsic contribution within the Kubo-Streda
formalism. In practice, one needs to evaluate Eqs.\,\eqref{eq:sigma1}-\eqref{eq:sigma2}
in the clean limit, which corresponds to the substitution for the
Green's functions $\mathcal{G}_{\mathbf{k}}^{a}\to G_{0\mathbf{k}}^{a}$.
Using the expression presented above for the latter, we can extract
the (single valley) $\sigma^{I}$ contribution as
\begin{equation}
\sigma_{yx}^{I}\left(\epsilon\right)=-\frac{1}{2}\frac{\lambda^{2}\delta}{\epsilon}\begin{cases}
\frac{2\left(3\lambda^{2}\delta^{2}+\delta^{4}-2\epsilon^{2}\left(\lambda^{2}+\delta^{2}\right)\right)}{\left(\lambda^{4}+3\lambda^{2}\delta^{2}+\delta^{4}-\epsilon^{2}\left(\lambda^{2}+\delta^{2}\right)\right)\sqrt{\epsilon^{2}\left(\lambda^{2}+\delta^{2}\right)-\lambda^{2}\delta^{2}}} & \text{region I}\\
\\
\frac{\left(2+\frac{\lambda}{\sqrt{\epsilon^{2}\left(\lambda^{2}+\delta^{2}\right)-\lambda^{2}\delta^{2}}}\right)}{\lambda^{2}+\delta^{2}+\sqrt{\epsilon^{2}\left(\lambda^{2}+\delta^{2}\right)-\lambda^{2}\delta^{2}}} & \text{region II}\\
\\
\frac{2\left(2\lambda^{2}+\delta^{2}\right)}{\lambda^{4}+3\lambda^{2}\delta^{2}+\delta^{4}-\epsilon^{2}\left(\lambda^{2}+\delta^{2}\right)}\,. & \text{region III}
\end{cases}
\end{equation}
For the Fermi sea term, $\sigma_{yx}^{II}$ we need instead to calculate
the derivative of the Green's functions with respect to the energy
variable. By doing that, performing the trace over internal indices
and angular integration we arrive at

\begin{equation}
\sigma_{yx}^{II}(\epsilon)=64\,\lambda^{2}\delta\lim_{\eta\to0}\text{Im}\left[\int_{-\infty}^{\epsilon}d\omega\int_{0}^{\infty}\frac{dk\,k}{2\pi}\frac{v^{2}k^{2}-\omega^{2}}{(v^{2}k^{2}-z_{1}^{R})^{2}(v^{2}k^{2}-z_{2}^{R})^{2}}\right]\,.\label{eq:SigmaII}
\end{equation}

\begin{figure}
\centering{}\includegraphics[width=1\columnwidth]{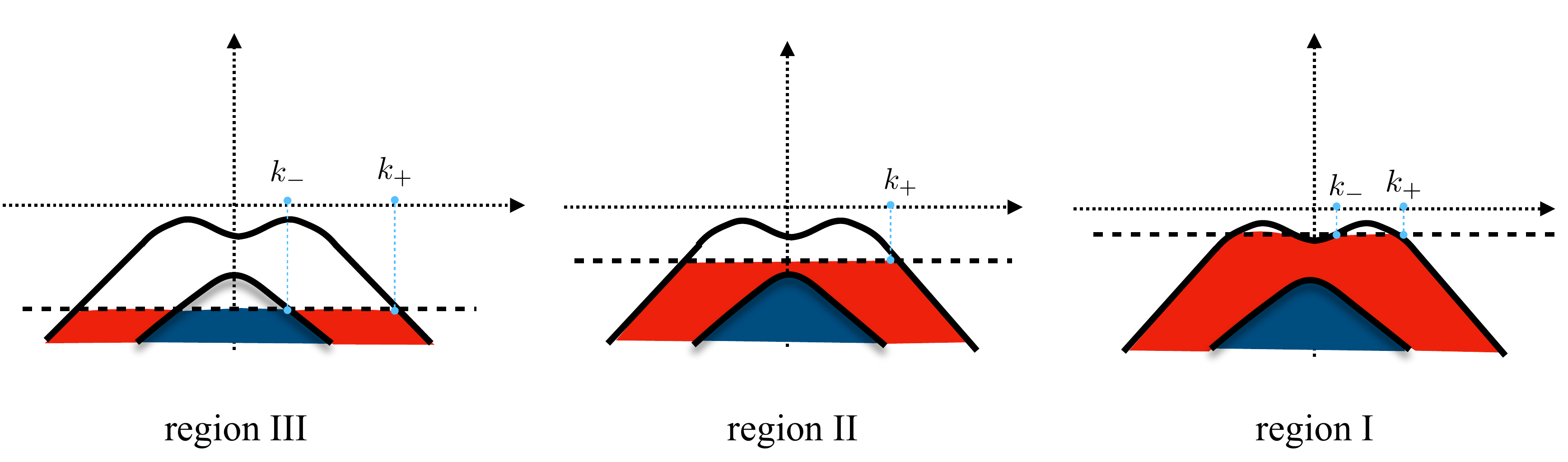}\caption{Schematic of band structure in the hole sector and different regimes\label{fig:Schematic-of-band}.
Depending on the position of the Fermi level $\epsilon$ the integration
over momentum variables has to be performed between different boundaries
as reported in details in Table \ref{tab:Summarising-table}. }
\end{figure}

Starting form Eq.\,\eqref{eq:SigmaII} we want to perform integration
over energies $\int_{\omega}$ first. Doing partial fraction decomposition
of the integrand 
\begin{equation}
g(k,\omega)=\frac{64\,\lambda^{2}\delta\,(v^{2}k^{2}-\omega^{2})}{(v^{2}k^{2}-z_{+})^{2}(v^{2}k^{2}-z_{-})^{2}}\,,\label{eq:integrand}
\end{equation}
we look for the poles in $\omega$. Obviously we get the eigenvalues
\begin{equation}
\epsilon_{\mu\nu}(k)=\mu\sqrt{v^{2}k^{2}+2\lambda^{2}+\delta^{2}+\nu\,S_{k}}\,,\qquad S_{k}=\sqrt{v^{2}k^{2}\left(\lambda^{2}+\delta^{2}\right)+\lambda^{4}}\,.
\end{equation}
displaced by the small imaginary part $i\eta$. Note the following
relations hold for the eigenvalues
\begin{align}
\epsilon_{++} & =-\epsilon_{+-}\,,\\
\epsilon_{--} & =-\epsilon_{-+}\,.
\end{align}
Let us write the partial fraction decomposition in the form
\begin{align}
g(k,\omega) & =\frac{A\omega+B}{\left(\omega-\epsilon_{-1}+i\eta\right)^{2}}+\frac{C\omega+D}{\left(\omega-\epsilon_{-2}+i\eta\right)^{2}}+\frac{E\omega+F}{\left(\omega+\epsilon_{-2}+i\eta\right)^{2}}+\frac{G\omega+H}{\left(\omega+\epsilon_{-1}+i\eta\right)^{2}}\,,\label{eq:partialFraction}
\end{align}
where we have labeled the eigenvalues according to the prescription
for the contracted index $n$ adopted in the main text
\begin{align}
\epsilon_{-1} & =\epsilon_{+-}\\
\epsilon_{-2} & =\epsilon_{--}\,,
\end{align}
which are respectively the lower and the upper branches of the spectrum
on the hole side (see Fig. \ref{fig:Schematic-of-band}). 
The eigenvalues appear as poles of second order. As no small $\eta$
is left in the denominator of the $\omega$-independent coefficients
$A,B,C,...$, we can focus on the imaginary part of the $\omega$
integration. We basically need to solve two classes of integrals
\begin{equation}
J_{a,b}=\int_{-\infty}^{\epsilon}d\omega\frac{\left(1,\omega\right)}{\left(\omega-\epsilon_{n}+i\eta\right)^{2}}\,.
\end{equation}
We have 
\begin{align}
J_{a} & =\left.\frac{-1}{\omega-\epsilon_{n}+i\eta}\right|_{-\infty}^{\epsilon}=\frac{-1}{\epsilon-\epsilon_{n}+i\eta}=-\frac{P}{\epsilon-\epsilon_{n}}+i\pi\delta\left(\epsilon-\epsilon_{n}\right)\,,\label{eq:Integral1}\\
J_{b} & =\int_{-\Lambda}^{\epsilon}d\omega\frac{\omega-\epsilon_{n}+i\eta}{\left(\omega-\epsilon_{n}+i\eta\right)^{2}}+\epsilon_{n}\,J_{a}=\log\left|\frac{\epsilon-\epsilon_{n}}{\Lambda}\right|-i\pi\theta\left(\epsilon-\epsilon_{n}\right)+\epsilon_{n}\,J_{a}\,,\label{eq:Integral2}
\end{align}
where in the last integral we have considered a regularization UV
cutoff $\Lambda$. When taking the imaginary part we have 
\begin{align}
J_{a} & =i\pi\sum_{\chi=\pm1}\frac{\delta\left(k-k_{\chi}\right)}{v_{\chi}}\,,\\
J_{b} & =-i\pi\left(\theta\left(\epsilon-\epsilon_{n}\right)-\epsilon\sum_{\chi=\pm1}\frac{\delta\left(k-k_{\chi}\right)}{v_{\chi}}\right)\,,
\end{align}
where we have used $v_{k_{\chi}}\equiv v_{\chi}$. The structure of
the integrals Eq.\,\eqref{eq:Integral1}-\eqref{eq:Integral2} is
important and allows to read already at this stage the quantization
of the transverse conductivity. To illustrate that, let us take negative
energies, where electronic states can populate bands labeled with
$-1,-2$. We can see now how the terms proportional to $\theta,\delta$
in Eqs.~\eqref{eq:Integral1}-\eqref{eq:Integral2} will contribute
differently depending on the position of $\epsilon$. In particular
we have the following table
\begin{table}[H]
\begin{centering}
$\int_{0}^{\infty}dk\,k\,g(k)\,\,$%
\begin{tabular}{|c|c|c|c|c|c|}
\hline 
 &  & I & II & III & gap\tabularnewline
\hline 
\hline 
$\theta\left(\epsilon-\epsilon_{-1}\right)$ & = & $\int_{k_{-}}^{\infty}dk\,k\,g(k)$ & $\int_{0}^{\infty}dk\,k\,g(k)$ & $\int_{0}^{\infty}dk\,k\,g(k)$ & $\int_{0}^{\infty}dk\,k\,g(k)$\tabularnewline
\hline 
$\theta\left(\epsilon-\epsilon_{-2}\right)$ & = & $\int_{k_{+}}^{\infty}dk\,k\,g(k)$ & $\int_{k_{+}}^{\infty}dk\,k\,g(k)$ & $\int_{0}^{k_{-}}dk\,k\,g(k)+\int_{k_{+}}^{\infty}dk\,k\,g(k)$ & $\int_{0}^{\infty}dk\,k\,g(k)$\tabularnewline
\hline 
 &  &  &  &  & \tabularnewline
\hline 
$\delta\left(\epsilon-\epsilon_{-1}\right)$ & = & $k_{-}g(k_{-})/v_{-}$ & 0 & 0 & 0\tabularnewline
\hline 
$\delta\left(\epsilon-\epsilon_{-2}\right)$ & = & $k_{+}g(k_{+})$ & $k_{+}g(k_{+})$ & $k_{-}g(k_{-})+k_{-}g(k_{-})$ & 0\tabularnewline
\hline 
\end{tabular}
\par\end{centering}
\caption{Table summarizing the momentum integrals in the different regimes\label{tab:Summarising-table}.}
\end{table}
The relevant observation is that inside the gap the $\delta$-parts
do not contribute, as $\epsilon$ does not intersect any band. Only
$\theta$-parts are left, with the integration extending from $0$
to $\infty$. According to the partial fraction decomposition of Eq.\,\eqref{eq:partialFraction},
we see, by using Eqs.\,\eqref{eq:SigmaII}, \eqref{eq:Integral2},
we are left with need to calculate 
\begin{equation}
\sigma_{yx}^{II}(\epsilon|_{\text{gap}})=\frac{1}{2\pi}\int_{0}^{\infty}dk\,k\,\left(A+C\right)\,.
\end{equation}
In this respect we can identify $A,C\equiv A_{\mathbf{k}},C_{\mathbf{k}}$
as the Berry curvatures of the lower and upper band respectively.
We discuss below how they exactly match a direct calculation of the
respective Berry curvatures. The analytic expression for the sum $A+C$
is quite complicated. However, explicit calculation shows (re-establishing
explicitly the units and considering the contribution of the $K^{\prime}$
valley)

\begin{equation}
\sigma_{yx}^{II}(\epsilon|_{\text{gap}})=\frac{2e^{2}}{h}.
\end{equation}
If the Fermi level lies instead outside the gap, $\sigma_{yx}^{II}$
acquires a energy dependence due to non-quantized $\theta$-parts
and a finite contribution of the $\delta$-parts. However, we verified
the $\delta$-parts cancel \emph{exactly} opposite in sign contributions
in $\sigma_{yx}^{I}(\epsilon)$. A similar cancelation has been reported
for a massive Dirac band model in Ref.\,\cite{supp4_Sinitsyn}\@.
We conclude the intrinsic contribution is a result of the $\theta$-parts
only. Finally note that for positive Fermi we can simply extract the
result from what we have obtained in the hole sector; we have in fact
$\sigma_{yx}^{\text{int}}(\epsilon)=\sigma_{yx}^{\text{int}}(-\epsilon)$,
which is a direct consequence of the Berry curvatures of the bands
summing to zero.

\subsubsection{Berry-curvature calculation}

We provide details now on a direct calculation of the intrinsic contribution
via the Thouless-Kohmoto-Nightingale-Nijs formula \cite{TKNNFormula}
\begin{equation}
\sigma_{yx}^{\text{int}}(\epsilon)=\sum_{n}\sum_{\mathbf{k}}\Omega_{\mathbf{k}}^{n}\,f^{0}(\epsilon_{n}(k))\,,
\end{equation}
where the band-BC is $\Omega_{\mathbf{k}}^{n}=\left(\nabla_{\mathbf{k}}\times\boldsymbol{\mathcal{A}}_{\mathbf{k}}^{n}\right)\cdot\hat{z}$
stemming from the Berry-connection $\boldsymbol{\mathcal{A}}_{\mathbf{k}}^{n}=-i\langle n\mathbf{k}|\nabla_{\mathbf{k}}|n\mathbf{k}\rangle$.
Let us focus again on the two bands in the hole sector. Performing
the calculation in polar coordinates, we find the Berry-connection
only contains the angular components:
\begin{align}
\left(\boldsymbol{\mathcal{A}}_{\mathbf{k}}^{n=-1}\right)_{\phi} & =\frac{\lambda^{2}\delta\left(\delta^{4}+2\lambda^{2}(2v^{2}k^{2}+3S_{k}+3\lambda^{2})+\delta^{2}(3v^{2}k^{2}+3S_{k}+4\lambda^{2})\right)}{\left(S_{k}^{2}\left(v^{2}k^{2}+2S_{k}+\delta^{2}+2\lambda^{2}\right)\right)^{3/2}}\,\,,\\
\left(\boldsymbol{\mathcal{A}}_{\mathbf{k}}^{n=-2}\right)_{\phi} & =-\frac{\lambda^{2}\delta\left(\delta^{4}+2\lambda^{2}(2v^{2}k^{2}-3S_{k}+3\lambda^{2})+\delta^{2}(3v^{2}k^{2}-3S_{k}+4\lambda^{2})\right)}{\left(S_{k}^{2}\left(v^{2}k^{2}+2S_{k}+\delta^{2}+2\lambda^{2}\right)\right)^{3/2}}\,,
\end{align}
with $S_{k}=\sqrt{\lambda^{4}+v^{2}k^{2}(\lambda^{2}+\delta^{2})}$.
We verified the latter expressions exactly agree respectively with
the coefficients $A,C$ of the previous section. It follows the results
match those from the Kubo-Streda formalism presented earlier.

\newpage

\begin{thebibliography}{10}
\bibitem{CommentIndex}We assign the label $n$ in descending order,
starting from the band of highest energy. Thus, $\nu,\mu=\pm(1,1)\to n=\pm2$
and $\nu,\mu=\pm(1,-1)\to n=\pm1$.
\bibitem{Comment1}We have exploited the fact that the bands of interest
are isotropic.
\bibitem{Comment2}The superscripts $I,\,II$ (italic) are not to
be confused with the indices $\text{I,\,II}$ (roman) indicating the
different regimes in the band structure, as depicted in Fig.\,1 of
the main text.
\end{thebibliography}
\end{document}